\documentclass[useAMS,usenatbib]{mn2e}
\usepackage{epsfig}
\usepackage{url}
\newcommand{\msun}{M_{\odot}}

\title[Parametrizing the Stellar Haloes of Galaxies]{Parametrizing the Stellar Haloes of Galaxies}

\author[R. D'Souza, G. Kauffmann, J. Wang, S. Vegetti]
{Richard D'Souza$^{1}$\thanks{E-mail address: rdsouza@mpa-garching.mpg.de
(RDS)},Guinevere Kauffman$^{1}$, Jing Wang$^{1}$, Simona Vegetti$^{1}$\\
$^{1}$Max Plank Institute for Astrophysics, Munich, Germany\\
}

\begin{document}

\date{Accepted 1988 December 15. Received 1988 December 14; in original form 1988 October 11}

\pagerange{\pageref{firstpage}--\pageref{lastpage}} \pubyear{2002}

\maketitle

\label{firstpage}

\begin{abstract}
We study the stellar haloes of galaxies out to 70-100 kpc as a function of stellar mass and galaxy type
by stacking aligned $r$ and $g$ band images from a sample of 45508 galaxies from SDSS DR9 in the redshift range $0.06\,\le\,z\,\le\,0.1$ and in the mass range 
$10^{10.0} \msun < M_{*} < 10^{11.4} \msun$r. We derive surface brightness profiles 
to a depth of almost $\mu_r \sim 32 \,\mathrm{mag\,arcsec}^{-2}$. We find  that the ellipticity of the
stellar halo is a function of galaxy stellar mass and that the haloes of
high concentration ($C > 2.6$) galaxies are more elliptical than those of
low concentration ($C < 2.6$) galaxies. The $g$-$r$ colour profile of high concentration galaxies reveals that the
$g$-$r$ colour of the stellar population in the stellar halo is bluer than in the main galaxy, and the colour of the stellar halo is redder for higher mass 
galaxies. We further demonstrate that the full two-dimensional surface intensity distribution of our 
galaxy stacks can only be fit through multi-component S\'{e}rsic models. Double-S\'{e}rsic profiles
adequately  model the average surface brightness distributions of high
concentration galaxies, while triple-S\'{e}rsic profiles are often
needed to model the surface brightness distributions  of low concentration
galaxies. Using the fraction of light in the outer component of the models 
as a proxy for the fraction of accreted stellar light, we show that this fraction is a function of stellar mass and galaxy 
type. For high concentration galaxies, the fraction of accreted stellar light rises from $30\%$ to $70\%$ for galaxies in
the stellar mass range from $10^{10.0} \msun$ to  $10^{11.4} \msun$. The fraction of
accreted light is much smaller in low concentration systems, increasing from $2\%$ to $25\%$ over the same mass range.
This work provides important constraints for the theoretical understanding of the formation of stellar haloes of galaxies.
\end{abstract}


\section{Introduction}
Traditionally, galaxies have been studied through their surface brightness profiles \citep{Hubble,deVauc}. 
This has not only revealed a wealth of information about their different morphologies but also hints about 
their formation processes. De Vaucouleurs (1948) first characterised the surface brightness profiles of giant elliptical 
galaxies as a simple $\log I(R) \propto R^{1/4}$ law, which was later also found to fit the bulges of disk galaxies. 
On the other hand, the disks of spiral galaxies have been traditionally fit with exponential profiles \citep{Freeman}. 
\cite{Sersic} showed that all these profiles are specific cases of a more general $\log I(R) \propto R^{1/n}$ 
function, which fits the surface brightness profile of a large number of galaxies from disks to spheroidals, 
dwarfs, ellipticals and bulges. The shape of the surface brightness profile provides valuable clues about the
way in which different galaxies formed.

As deeper and more resolved surface brightness data became available, deviations from these 
simple laws became clearly evident, indicating that galaxy formation was a more complex
process than previously believed \citep{Kormendy}. 
This discovery  motivated the use of multiple components to model the surface brightness profiles of galaxies 
\citep{Kormendy77, Simard, Lackner}  
Bulge-disk decompositions helped distinguish pseudo-bulges ($n\sim 1$) from classical bulges ($n \sim 4$). 
Pseudo-bulges are dense central components of disk galaxies that are flattened and 
rotationally supported and believed to be built out of disk gas. Classical bulges lie on the fundamental
plane linking galaxy size, luminosity and velocity dispersion \citep{Bender}.

With the advent of deeper imaging (through Hubble Space Telescope and medium-sized, ground-based telescopes), it has become 
possible to detect additional fainter stellar structures around both
normal galaxies and brightest cluster galaxies \citep{Schweizer_a, Malin, Schweizer_b} 
Today, stellar haloes of galaxies have been observed and confirmed not only in clusters 
as intracluster light (ICL), but also 
in a large variety of field galaxies from early-type to late-type spirals. This is consistent with the idea that
the faint stellar halo is built up from the debris of smaller galaxies and satellites that are tidally 
disrupted (e.g. \citealt{Bullock} and \citealt{Cooper10}).

In the Milky Way and in other nearby disk galaxies, the stellar halo and other tidal features have been directly
detected through star counts \citep{Bell2,Ibata,Monachesi}. Observing the stellar halo through star counts 
is limited to the Local Universe. The integrated light from deep imaging has enabled studies of
the stellar haloes of more distant elliptical and spiral galaxies (see e.g. \citealt{mihos}, \citealt{delgado}, 
\citealt{Tal2} and \citealt{dokkum}). By using modest aperture telescopes \citep{delgado} 
with innovative telescope design optimised for low surface brightness emission \citep{dokkum}, one can reduce the 
systematic errors related to flat fielding and the complex point spread function (PSFs) 
of the telescope and reach much 
deeper limiting depths of $\mu_g \sim 32 \,\mathrm{mag\,arcsec}^{-2}$.

Alternatively, stacking the images of a large number of similar galaxies (e.g. \citealt{Zibetti04}, 
\citealt{Zibetti05}, \citealt{Tal} and \citealt{Cooper}) enables one to study the average stellar haloes of 
statistical samples of more distant galaxies. The disadvantage is that information on detailed
structure is lost. \cite{Zibetti04} used stacking techniques to study the stellar 
haloes of edge-on disk galaxies; \cite{Tal} studied the stellar haloes of luminous red galaxies out to $z \sim 0.34$.

Theoretical models \citep{Cooper,Purcell,Oser,Lackner2} predict not 
only large variations between individual stellar haloes of galaxies, 
but also systematic variations in the average
properties of stellar haloes as a function of certain galaxy parameters (for example, halo mass, stellar mass, galaxy bulge-to-disk ratio, etc). In order to constrain theoretical models for the formation of stellar haloes, it is important 
to study the average properties of the surface brightness profiles of galaxies as a function of these galaxy parameters.
In this paper, we stack a large number of galaxy images and study 
them as a function of stellar mass and galaxy type (late-type or early-type). 
The SDSS imaging data set is well-suited to study the faint 
stellar haloes of galaxies \citep{Zibetti04,Zibetti05,Tal}. The systematics 
of stacking many SDSS images to produce a very deep image have been well 
understood and quantified. This is important because studying low-surface brightness
structures is highly dependent on a proper estimation and removal of the sky background. We pay particular attention to the 
residual sky background obtained after stacking the sky-subtracted images from SDSS DR9.
We then model the surface brightness profile of the stacked galaxy including the stellar halo through multi-component fits. 
We then parametrise the contribution of the stellar halo by deriving the fraction of light in the outer
component of the galaxy. 

In Section \ref{sec:selection}, we describe how we select and prepare our galaxy images for 
stacking. In Section \ref{sec:stacking}, 
we describe in detail the stacking procedure, our error analysis, PSF analysis and the 
methodology we employ to derive the ellipticity, surface brightness and the colour profiles
for each galaxy stack. In Section \ref{sec:analysis}, we present the surface brightness and 
colour as a function of the stellar mass of the galaxy and of galaxy type. In Section \ref{sec:modelling}, 
we fit models to these surface brightness profiles and determine the 
fraction of light in the outer faint stellar component. In Section \ref{sec:summary}, we summarise 
and in Section \ref{sec:discussion}, we discuss our results in light of our theoretical understanding of the formation of
stellar haloes of galaxies. Throughout this paper, we assume a flat $\Lambda$CDM 
cosmology, $\Omega_{\mathrm{m}}=0.25$, $\Omega_{\mathrm{\Lambda}}=0.75$ and Hubble parameter $h=0.73$.

\section{Sample Selection and Image Preparation}
\label{sec:selection}
We select isolated central galaxies from the MPA-JHU SDSS spectroscopic `value-added' catalogue in the stellar mass 
range $10^{10.0} \msun < M_{*} < 10^{11.4} \msun$ and in the redshift range $0.06\,\le\,z\le\,0.1$.\footnote{The stellar
masses used here are as defined by the MPA-JHU catalogue (using a methodology similar to 
that described in \citealt{Kauffmann03a}) and 
corrected for the Hubble parameter $h=0.73$. The stellar mass estimates in the MPA-JHU catalogue were derived from fits to the SDSS 
fibre photometry and the total \texttt{ModelMag} photometry.} We apply the isolation criterion outlined in  \cite{Wang}: 
a galaxy of apparent $r$-band magnitude $m_{central}$ is considered isolated 
if there are no galaxies in the spectroscopic catalogue 
at a projected radius $R < 0.5\, \mathrm{Mpc}$ and velocity offset  $|\delta z| < 1000  \, 
\mathrm{km \,s}^{−-1}$ with magnitude 
$m < m_{central} + 1$, and none within $R < 1 \, \mathrm{Mpc}$ and $|\delta z| < 1000  \, \mathrm{km\, s}^{−-1}$ with 
$m < m_{central}$. We remove all edge-on disk galaxies to avoid adverse PSF effects along the minor axis \citep{dejong}
by choosing only those galaxies with isophotal minor-to-major axis ratio $b/a > 0.3$.

We construct mosaics (1200 x 1200 pixels) in the $g$, $r$ and $i$ bands centred on each galaxy using the  
sky-subtracted SDSS Data Release 9 images and \texttt{SWarp} \citep{Bertins2}. Galaxies were removed if
found unsuitable for stacking. First, galaxy images with a bright source with an $r$-band petrosian 
magnitude greater than 12.0 and within a distance of 1 Mpc from the centre of the 
galaxy were removed. Secondly, if the masking algorithm (outlined later) failed due to crowded fields, the  galaxy
image was discarded. Finally, we calculated a histogram of the difference between each galaxy mosaic 
after masking and transformation (see
later) and the stacked image. Galaxy mosaics lying more than $5\,\sigma$ from the mean were discarded.
The final sample contains a total of 45508 galaxies.

\begin{figure}
  \includegraphics[width = \linewidth]{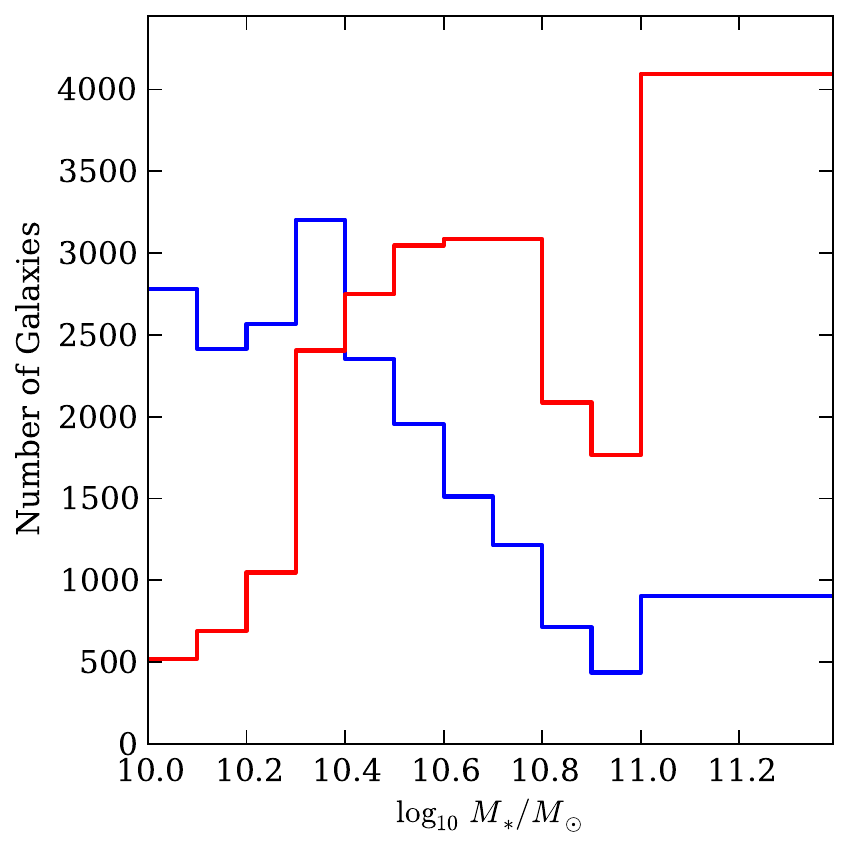}
  \caption{The number of galaxies in each mass bin (0.1 dex in width) split according to low concentration ($C<2.6$ blue line) and high concentration ($C>2.6$
    red line) galaxies. The highest mass bin is 0.4 dex in width.
  }
  \label{fig:histogram}
\end{figure}

For our later analysis, we will stack according to stellar mass and concentration. 
For the stellar mass stacks, we stack galaxies
in stellar mass bins of 0.1 dex. For the highest mass bin we stack in a bin size of 0.4 dex. Each stack contains both early and 
late-type galaxies: late-type galaxies dominate the stacks of lower stellar mass  whereas early-type galaxies are predominant at high stellar masses. We can parametrise the shape of the galaxy by using the concentration index $C=R90/R50$ (where 
$R90$ and $R50$ are the radii containing 90 and 50 per cent of the Petrosian $r$-band luminosity of the galaxy). It has been 
demonstrated that $C \sim 2.6$ marks the transition from late-type to early-type morphologies \citep{Strateva}. 
In order to study the stellar halo separately for late-type and early-type galaxy morphologies, we divide our sample into stellar 
mass bins of 0.2 dex with a further separation of each stack into high concentration ($C>2.6$) and low concentration galaxies ($C<2.6$). 
The number of galaxies in each stack is displayed visually in Figure \ref{fig:histogram}.

Conservative masking was employed by using multiple runs of \texttt{SExtractor} \citep{Bertins1} to create segmentation maps. 
For this purpose, the mosaics of three bands were stacked together to make a `master image', from which several segmentation 
masks were created to deal with various types of background and overlapping objects. We used a minimum detection area of 5 pixels,
a Gaussian filter for detection and a detection threshold of $1.5\,\sigma$ to create all the masks. For the background detection, we use three variations. 
We first calculated the mask with a global background. We then calculate the mask with a local background size of 256 pixels with a filter 
of 20 pixels. Later we calculated a mask with a smaller background size of 128 pixels with a similar filter size. To deal with extended 
faint objects, a mask was also created by convolving the master image with an $8\times8$ pixel top 
hat kernel before running \texttt{SExtractor}. 
Each of these masks were successively applied to individual $g$ and $r$-band mosaics. 
The $i$-band mosaics were only used for creating the 
master images for the masking procedure.

The masked mosaics were then transformed to $z=0.1$ with the flux-conserving IRAF task \texttt{GEOTRAN}. This involves both a cosmological surface
brightness dimming $(1+z)^4$ and an image rescaling. For the final transformed mosaic at $z=0.1$, $1\,\mathrm{pixel}\,=\,0.71\,\mathrm{kpc}$. The mosaics
were further cropped to a uniform size of $950\times950$~pixels ($550\times550$~kpc at $z\sim0.1$) and corrected for Galactic extinction following \cite{Schlegel}.
We ignored $K$-corrections in scaling the images as they tend to be minimal at $z<0.1$.

A sizable number of the final transformed images are oversampled. However, for the redshift shift range chosen for our sample $z=0.06-0.1$, this
does not significantly affect the noise characteristics of our final transformed images. A final run of SExtractor was used to determine the 
position angle of the galaxy in the $r$-band mosaic. This position angle is measured 
by calculating the second-order moments of the intensity
distribution and corresponds to surface brightness threshhold  $\mu_r \sim 24 \, \mathrm{mag\,arcsec}^{-2}$, 
or a radius of $\sim$10 kpc.  Each mosaic was then rotated using 
\texttt{GEOTRAN} such that the major axis of each galaxy was aligned. 

We note that combining galaxy images into mosaics may introduce additional systematics. \cite{Blanton} compared the mosaics 
created from the sky-subtracted images of DR9 and those created directly from the raw images and found that 
they yield equivalent results. 

The sky subtraction in DR9 \citep{Blanton} is a remarkable improvement from early data releases especially 
for the extended low surface brightness regions around low-redshift galaxies. \cite{Blanton}  calculate the 
residual sky background by measuring the mean surface brightness in random patches of size 13 x 13 native 
SDSS pixels marked as ``sky" in the SDSS pipeline across all  imaging runs (see Figure 5 of \citealt{Blanton}. 
These residuals become significant at depths beyond  $\mu_r \sim 26 \, \mathrm{mag\,arcsec}^{-2}$. We will 
discuss this further in the next section.

\section{Image Stacking and Methodology}
\label{sec:stacking}
\subsection{Stacking Procedure}
Each stack contains between 1000 and 5000 galaxies with an average of 3000 galaxies. The mosaic images in the $g$ and $r$ bands were stacked using the IRAF task \texttt{IMCOMBINE}, 
by taking the mean value of each pixel after clipping at the  10th and 90th percentiles.\footnote{Percentile clipping also helps prune any 
close satellite galaxies which escape the masking procedure.} The images were not weighted in the stacking process so as not to bias the sample. 
The masked parts of the images were not used when calculating the mean value in \texttt{IMCOMBINE}. To make the stacking computationally easier, 
the final stacks were built by combining equal stacks of around $\sim 100$ galaxy images each. By working in narrow mass bin ranges, we avoid the 
difficult problem of normalising the size of images in each bin prior to stacking.

\subsection{Estimation of Background for Stacked Galaxies}
The background ``sky'' for individual DR9 images consists of the 'residual' sky background and 
light from undetected (unmasked) galaxies. 
In the Appendix \ref{appendix:light}, we quantify the level of light from undetected  sources. This tends to be minimal due to the strict masking procedures 
employed and the fact that we only select isolated galaxies. 

To estimate the residual sky background for the stacked image, we calculate the mean intensity in an annulus 
between 280 and 320 kpc (400-450 pixels) from the centre of the stacked image. We assume that this background 
is constant over the whole image. To calculate the uncertainty in this background estimation, we calculate
the standard deviation of the mean calculated in patches of 16 x 16 pixels within this annulus.

\begin{figure}
  \includegraphics[width = \linewidth]{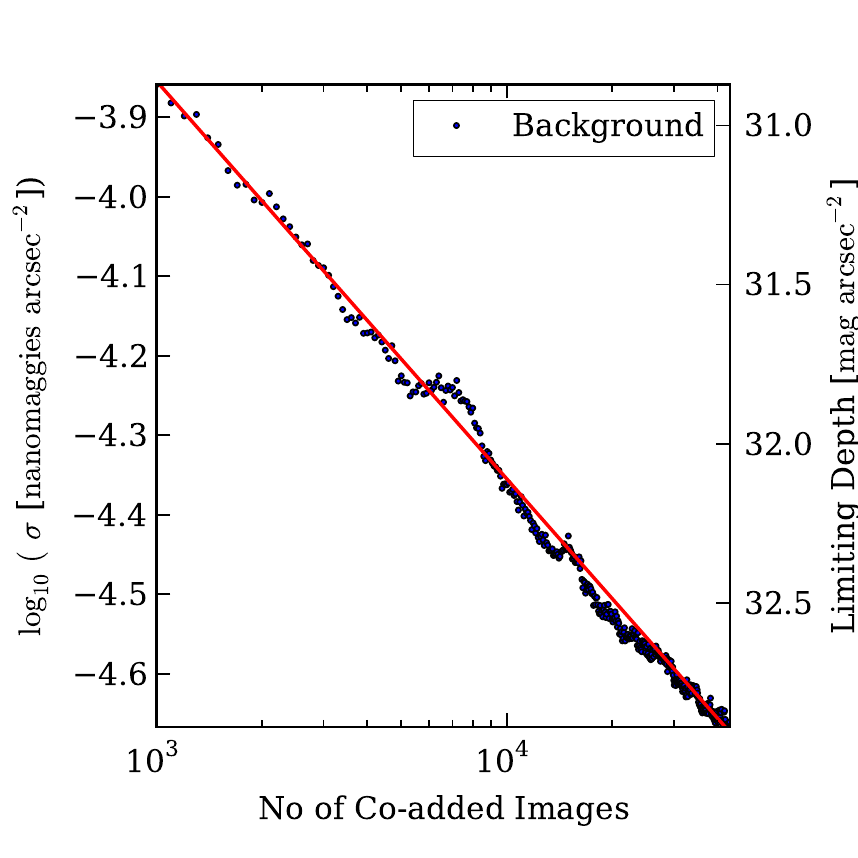}
  \caption{The logarithm of the uncertainty in background removal as a function of the number of co-added images. 
    The right axis depicts the limiting surface brightness depth. The red line indicates the function
    $0.00442/\sqrt N_{Images} \mathrm{nanomaggies\,arcsec}^{-2}$.
  }
  \label{fig:limiting_mag}
\end{figure}

With the standard SDSS imaging, it is possible to extract radial surface brightness profiles down to $\mu_r 
\sim 27 \, \mathrm{mag\,arcsec}^{-2}$ \citep{Pohlen}. With a better residual background estimation of high S/N stacked DR9 
images, it is possible to go significantly deeper. In Figure \ref{fig:limiting_mag}, we plot the uncertainty in 
the residual background estimation and the corresponding limiting depth in the $r$-band as a function of the 
number of co-added objects. The uncertainty in the residual background estimation can be fit by the function 
$0.00442/\sqrt N_{Images} \mathrm{nanomaggies\,arcsec}^{-2}$.

\subsection{Error Estimation}
For stacks of a few thousand galaxies, the formal uncertainty in the stacked surface brightness
profiles at larger radii is dominated by the uncertainty in subtracting the background sky, which
consists of camera noise plus extragalactic background radiation originating in the  
stellar populations of galaxies at moderate to high redshift. These uncertainties
calculated as described in the previous section are depicted as solid error bars in the plots 
discussed in the next section. In addition to the uncertainty that arises from the sky subtraction, 
it is interesting to consider the variance that arises from the fact that similar galaxies may have stellar
haloes with quite different masses and sizes. This can be quantified for each
pixel in our final $g$ and $r$ band stacks through 
a bootstrapping procedure. For each bin, 3000 stacks were created with repetition 
and the variance in each pixel is calculated for each band. This gives the total uncertainty of each 
pixel. After accounting for the formal uncertainty, the variance in the surface brightness profiles 
can be calculated and is depicted as shaded regions in the plots. 

To verify that the faint outer stellar halo visible in our stacks 
between $30-32\, \mathrm{mag\,arcsec}^{-2}$ is not a product of
systematics in the data or due to our stacking procedure, we created equivalent 
background stacks (nearly 3000 images) for each 
bin by choosing a location 5 Mpc away from the centre of the galaxy in a random direction where 
no large galaxies were found within a distance of 1 x 1 Mpc. 
We found that evaluating the background at these very large distances made no difference to our results.

\begin{figure}
\includegraphics[width = \linewidth]{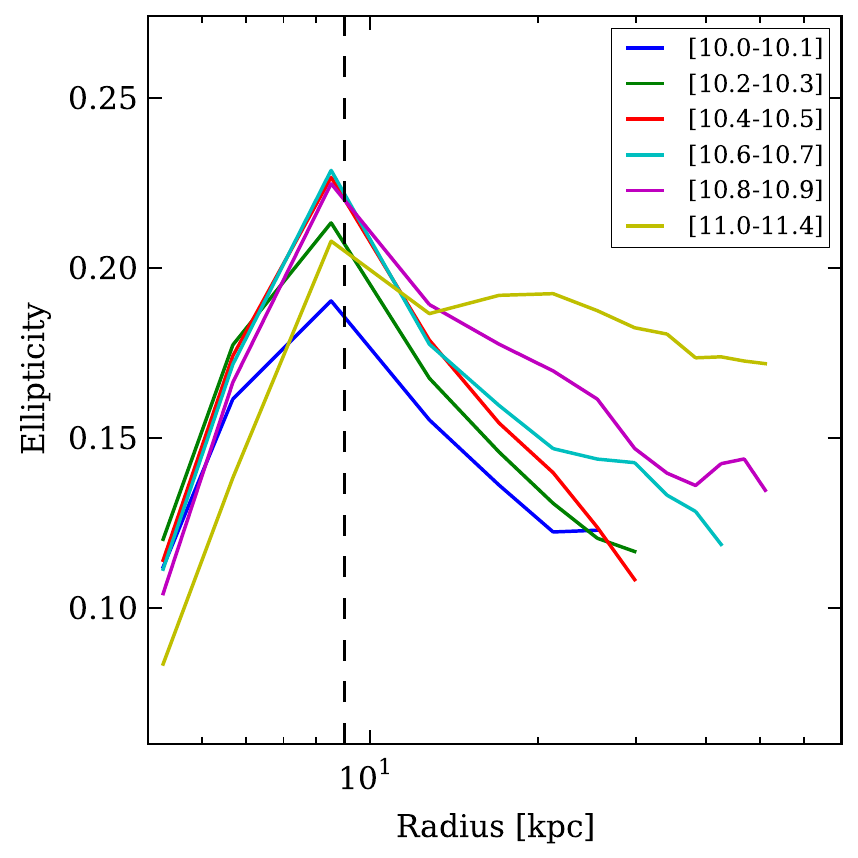}
\includegraphics[width = \linewidth]{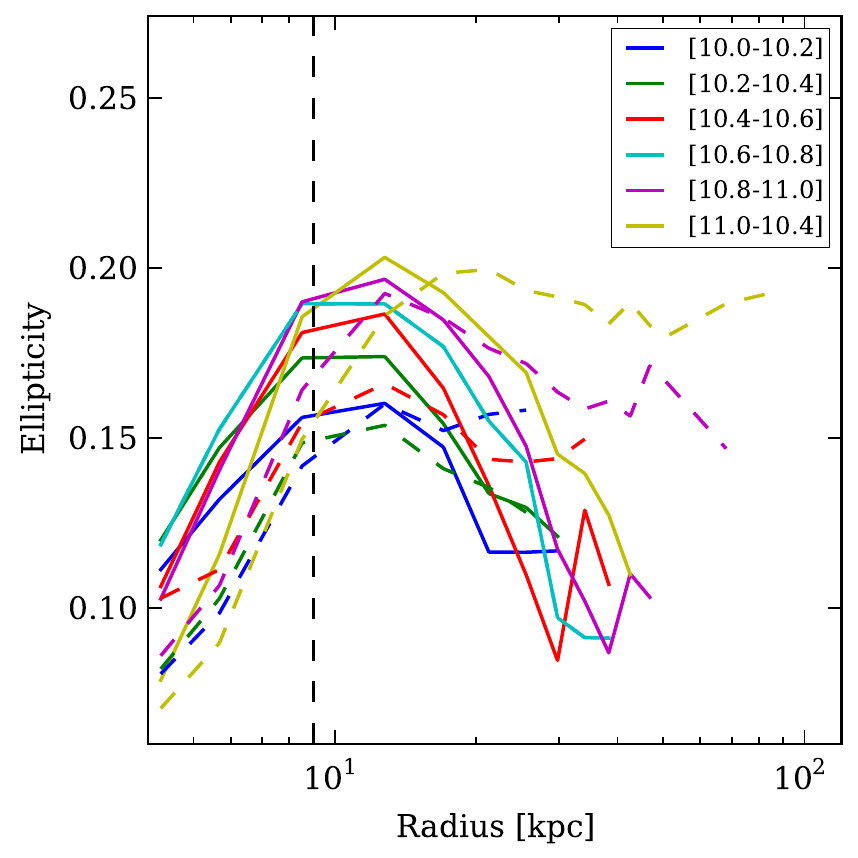}
\caption{(a) Ellipticity profiles for successive stellar mass bins. (b) Ellipticity profiles for each of the stellar mass bins divided 
according to concentration. Solid lines and dashed lines indicated low ($C<2.6$) and high ($C>2.6$) concentration galaxies respectively. 
The vertical dashed line indicates the maximum radius affected by the PSF.}
\label{fig:stellarmassbinellip}
\end{figure}

\begin{figure}
  \includegraphics[width = \linewidth]{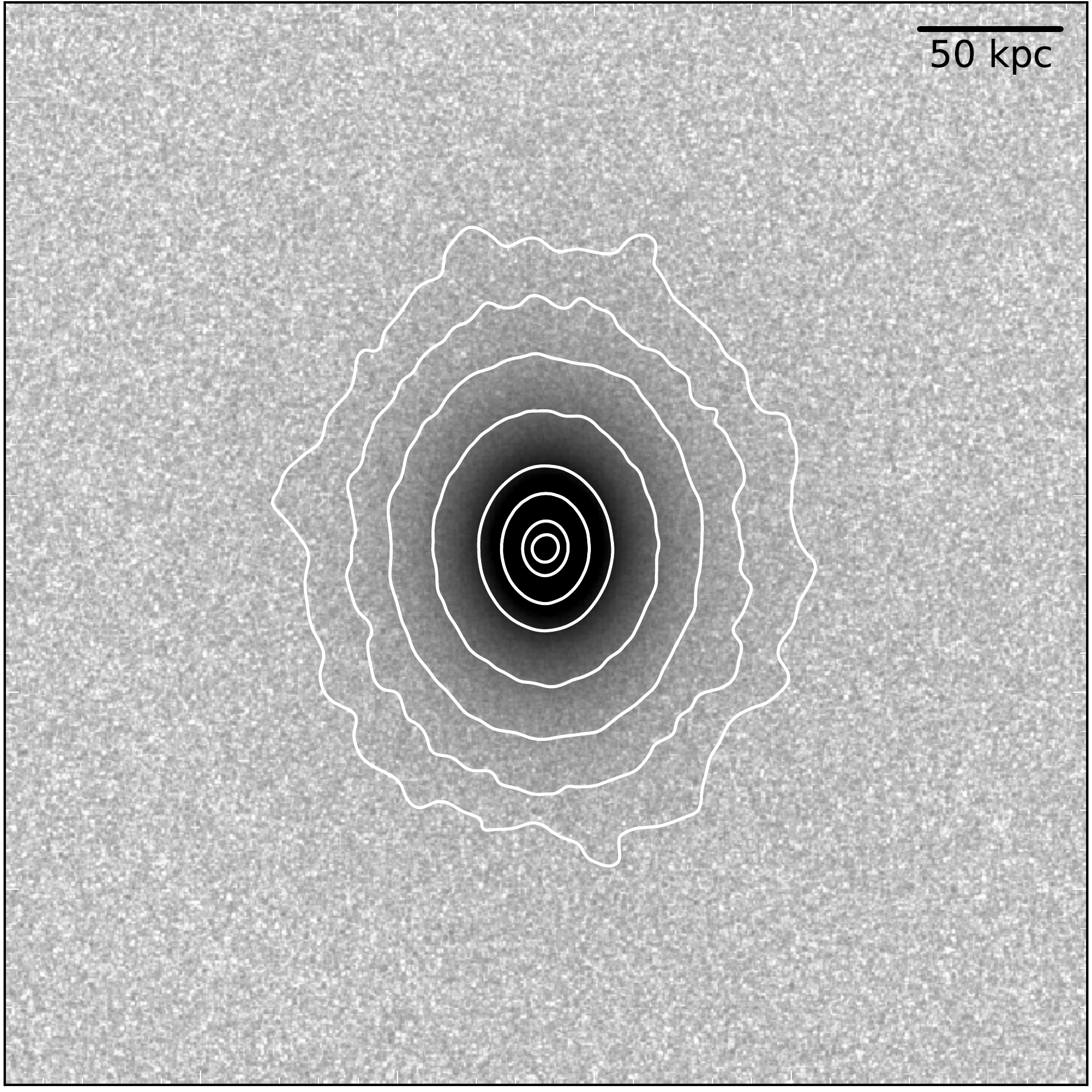}
  \caption{The stacked image consisting of 4040 images in the mass range $10^{11.0} \msun < M_{*} < 10^{11.4} \msun$ and $C>2.6$. 
    Elliptical contours are drawn at 5, 10, 20, 30, 50, 70, 90 and 110 kpc.}
  \label{fig:image_contours}
\end{figure}

\begin{figure}
  \includegraphics[width = \linewidth]{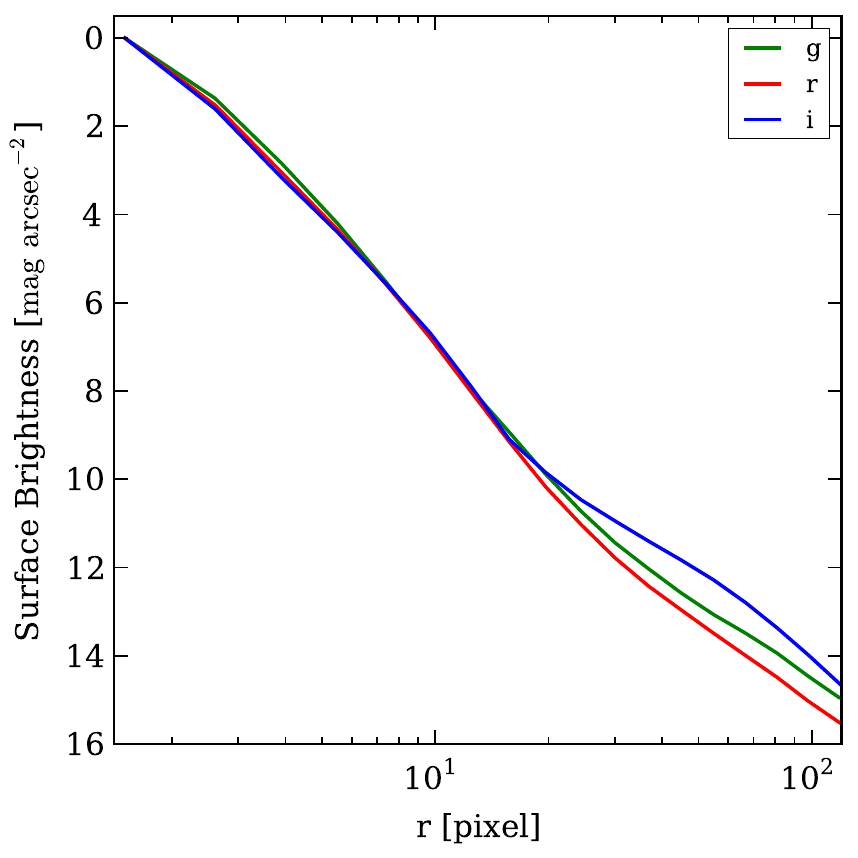}
  \caption{SDSS Point Spread Functions colour coded for the $g$, $r$ and $i$ bands as indicated by the legend.}
  \label{fig:psf}
\end{figure}

\subsection{PSF Effects}
\label{section:psf}
The PSF flattens the ellipticity and the surface brightness profiles at the centre of the galaxy 
at radii less than $\sim 10$ kpc.
For deep images, the light in the faint outskirts of the stack can be dominated by the scattered 
light from the centre of the galaxy. Failure to account for the difference in the extended wings 
of the PSF, especially in the $i$-band, can lead to a reddening of the colour of the stellar halo 
\citep{dejong}. This is very visible along the minor axis of edge-on disk galaxies where the 
surface brightness decreases faster than the profile of the wings of the PSF.

We choose not to deconvolve the stacked galaxy profiles. 
The effect of the PSF is much smaller in our work than that 
of \cite{Tal} due to the fact that the galaxies are much closer in redshift.
For data interpretation purposes, we
will model the two-dimensional stacked image of the galaxy convolved with the average PSF. 
We have thus constructed average PSF stacks in the $g$, $r$ and $i$ bands by combining the synthetic 
PSFs created using Robert Lupton's Read Atlas Images 
code\footnote{\url{http://www.sdss.org/DR7/products/images/read_psf.html}} 
and stacked bright star images according to the procedure outlined in \cite{Tal}. 
The PSF profiles for the $g$, $r$ and 
$i$ bands are shown in Figure \ref{fig:psf}. 

Due to the fact that the PSFs in the $g$ and in the $r-$ bands are
similar (see also Fig 2 of \citealt{dejong} as well as Fig 6 
of \citealt{Bergvall}), our $g$-$r$ colour profiles are not significantly 
affected by PSF effects, especially in the outer parts of the profiles.  
However, the 
$i$-band PSF does differ significantly (see Figure \ref{fig:psf}) in having
wings that extend to much larger distances. 
We therefore avoid the use of the SDSS $i$-band.

\subsection{Ellipticity, Surface Brightness and Colour Profiles}
Measuring the ellipticity can help quantify the shape of the average stellar halo. The ellipticity profiles ($1 - b/a$) for each of the 
aligned galaxy stacks are determined by generating intensity contours at various distances from the centre of the stacked image of the 
galaxies in the $r$-band. For deriving contours which were greater than 20 pixels away from the centre of the galaxy stack, we smooth 
the image with a Gaussian filter with a width of 3 pixels. For contours beyond 60 pixels from the centre of the galaxy stack, we smooth the 
image with a larger Gaussian filter (width of 5 pixels). 

In Figure \ref{fig:stellarmassbinellip}, we plot ellipticity profiles out to radii of  30-50 kpc 
for our stacks divided according to  stellar mass and concentration. 
Information on the shape of stellar haloes can be inferred from the average ellipticity profiles for each stack. 
Only the inner part ($ < 10\,\mathrm{kpc}$) of the ellipticity profile is significantly affected by the PSF. The outer parts of the ellipticity 
profile show a gradual change in ellipticity with radius. The ellipticity profile of the stacks of lower stellar mass decreases as the radius 
increases, i.e. for these galaxies the outer part of the stellar halo is more circular than the inner part of the galaxy. The ellipticity 
of the outer part of the stellar halo increases as a function of $M_*$. The highest stellar mass bins have a maximum outer 
ellipticity of $\sim 0.17$, which remains approximately constant from 30 to 50 kpc. 
In Figure \ref{fig:image_contours}, we show the stacked image of 
high concentration galaxies stacked in the mass range  
$10^{11.0} \msun < M_{*} < 10^{11.4} \msun$ along with elliptical contours drawn at various radii. 

We find that the stellar haloes of low concentration galaxies tend to be spherical, while the stellar haloes of high 
concentration galaxies tend to be elliptical. At fixed mass, the ellipticity
of the highest stellar mass, high concentration galaxies reaches values of  0.2 and is approximately constant
from 20 to 100 kpc.
By contrast, the measured ellipticity ($1-b/a$) of low concentration 
galaxies is around 0.1. 

Are these results consistent with other measurements? The stellar halo of M31 can easily be measured out to 
large distances and is found to be nearly spherical \citep{Ibata}.
At 80 kpc for high concentration high stellar mass galaxies, the measured 
ellipticity is $0.21 \pm 0.08$.  This is also consistent with the ellipticity of the 
stellar halo of LRGs measured by \cite{Tal} which lies around $\sim 0.25-0.3$. 
On the other hand, \cite{Sesar} measured the axial ratio of the Milky Way stellar 
halo out to a distance of 35 kpc and estimated it as  $q \sim 0.7$, i.e. an ellipticity of 0.3, which
lies outside the range spanned by our estimates. This may imply that the Milky Way's halo is unusual.
We note, however,  that when stacking aligned galaxies together, we assume
that the outer stellar halo is also aligned with the shape of the galaxy. If this were not the case, 
it would lead to a systematic uncertainty in the intrinsic ellipticity which
would increase with radius. As a result, the ellipticity measured is a lower 
limit on the true average intrinsic ellipticities of the stellar haloes of the galaxies which make up the stack. 
Convolving the stacked images creates additional measurement uncertainties. 

Using these ellipticity profiles, we derive surface brightness in the 
$r$ band and $g$-$r$ colour profiles in elliptical annuli after  
background subtraction. 
At radii where the ellipticity estimates are no longer reliable, we assume that the 
ellipticity profile flattens out at the furthermost determined value of the  ellipticity.

\section{Analysis of Stacked Images}
\label{sec:analysis}
\subsection{Profiles in Stellar Mass Bins}
\begin{figure*}
 \includegraphics[width = 0.49 \textwidth]{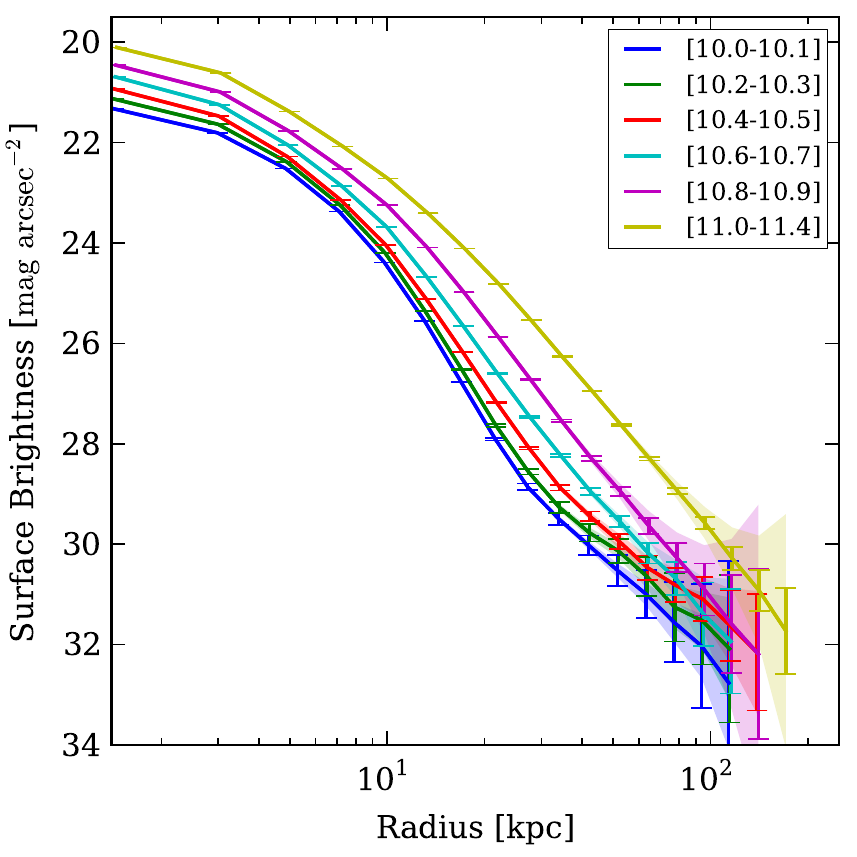} 
 \includegraphics[width = 0.49 \textwidth]{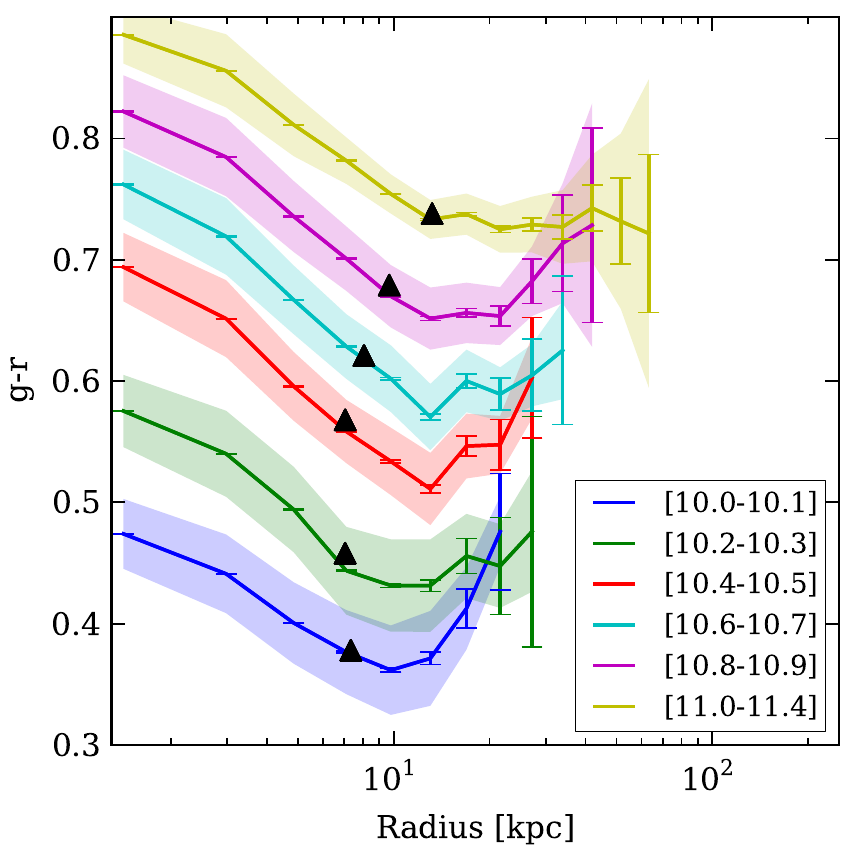} 
\caption{
  Surface brightness profiles and $g$-$r$ colour profiles of stacks for successive stellar mass bins.
  The error-bars show the sum of instrumental errors and uncertainty in background subtraction, 
while the shaded regions show the spread 
due to the variation in the shape of the stellar halo. The triangles in the colour profiles mark the average R50 of
galaxies in  the respective bin.
}
\label{fig:stellarmassbin}
\end{figure*}

In Figure \ref{fig:stellarmassbin}, we show the average surface brightness profiles and 
the average $g$-$r$ colour profiles for our galaxy stacks in stellar mass bins.
The surface brightness profiles extend reliably to a depth of  $\mu_r \sim 32 \quad\textrm{mag\,arcsec}^{-2}$.
The profiles of highest mass bins reach out to 100-150 kpc, while the lower mass bins extend up to 60-100 kpc. 
The surface brightness profile of the stellar halo show variations
with stellar mass. As discussed in \cite{Cooper}, the trend in the surface brightness 
profiles in the stellar mass range $10^{10.7} \msun < M_{*} < 10^{11.4} \msun$ 
is consistent with the theoretical predictions. In this paper, we extend the analysis 
down to $10^{10} M_{\odot}$; comparison with model predictions will form the
subject of a future paper.  

The triangle markers in the colour profiles indicate the average $R50$ 
(the radius enclosing 50 per cent  of the Petrosian $r$-band luminosity of the galaxy) 
for each mass bin. For each mass bin, there is a flattening in the colour profile 
and a hint of an upturn beyond the average $R50$ indicating that we may be  seeing 
the effects of an older accreted component. We will quantify this in more detail in the next section.

\subsection{Profiles in Stellar Mass Bins divided by Concentration}
\begin{figure*}
  \begin{tabular}{cc}
  \includegraphics[width = 0.49 \linewidth]{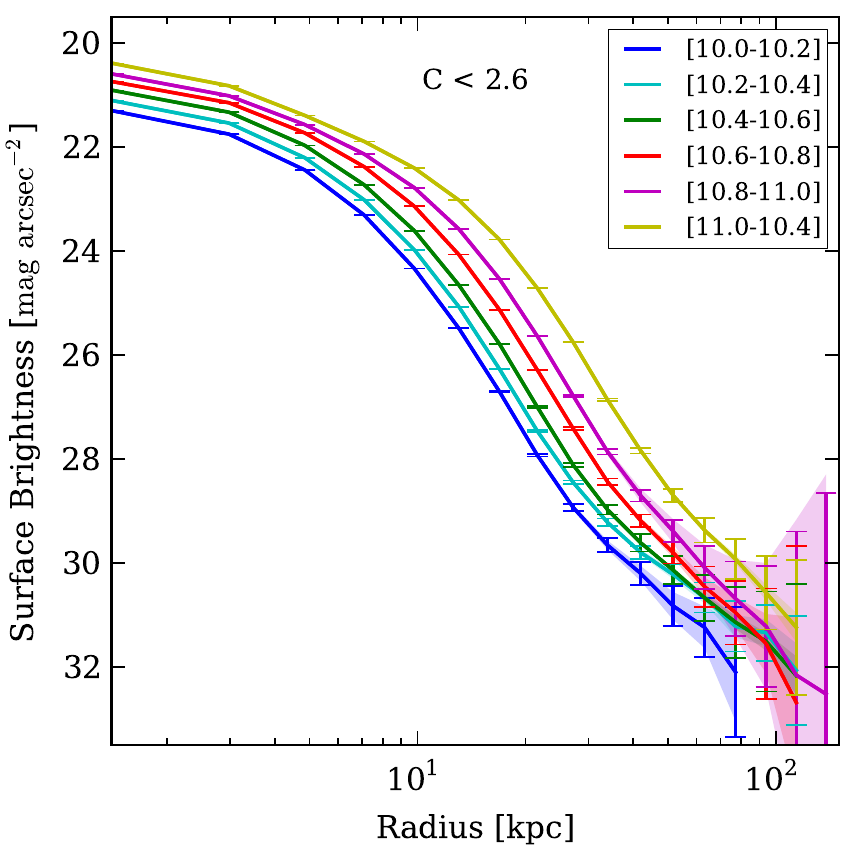} &
  \includegraphics[width = 0.49 \linewidth]{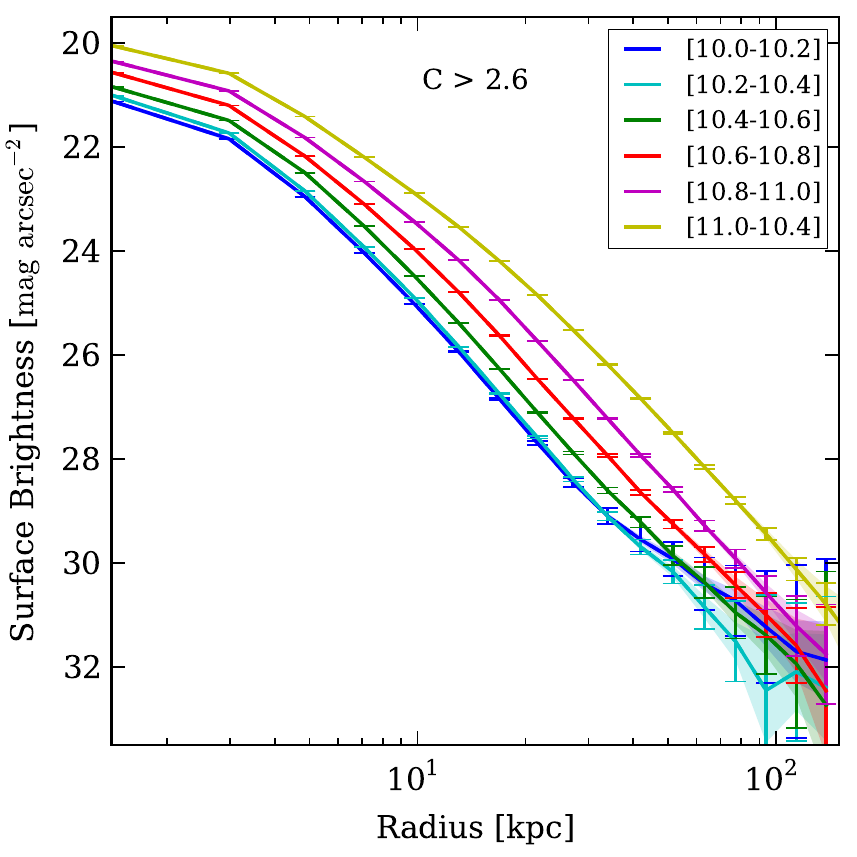} \\
  (a) & (b) \\
  \includegraphics[width = 0.49 \linewidth]{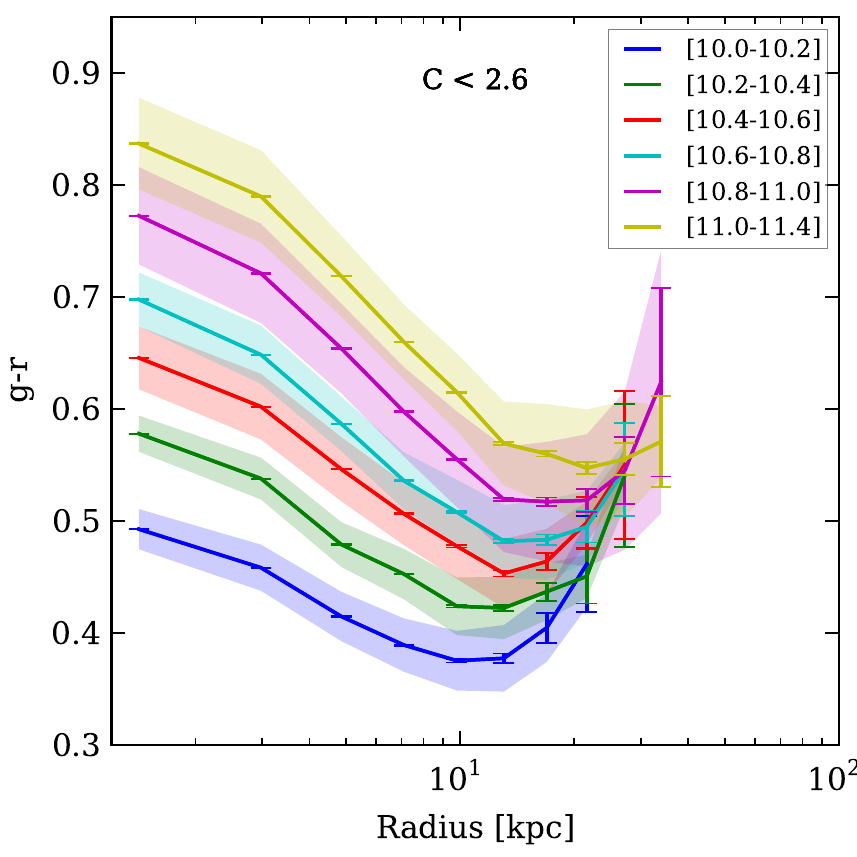} & 
  \includegraphics[width = 0.49 \linewidth]{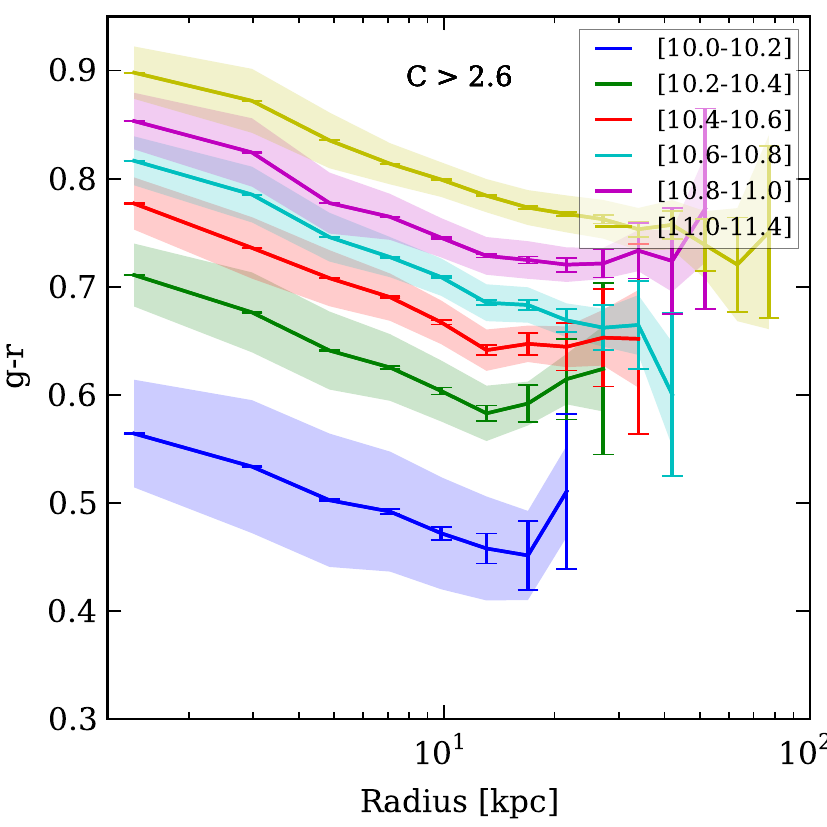} \\
  (c) & (d) \\
  \end{tabular}
  \caption{
    Surface brightness and $g$-$r$ colour profiles of the stellar mass bins divided according to concentration. 
    The error-bars show the sum of the instrumental errors 
    and the uncertainty in background subtraction, 
    while the shaded regions show the spread due to the variation in the shape of the stellar halo.
  }
\label{fig:highlowconc}
\end{figure*}

In Figure \ref{fig:highlowconc}, we show the average surface brightness profile and the average $g$-$r$ colour profiles for our galaxy stacks
separated into high ($C>2.6$) and low ($C<2.6$) concentration galaxies. 

The surface brightness profiles reveal a clear difference in the shapes 
of the stellar haloes of high concentration and low concentration galaxies. 
We can parametrise the shape of the stellar halo by measuring its outer slope.
The outer slope is measured through a Bayesian methodology that takes into 
consideration the scatter due to the variance of the shape of the surface 
brightness profile of the galaxy. The details are outlined in Appendix \ref{appendix:slope}.
  
In Figure \ref{fig:slope_lc_uc}, we plot the 
slope $\Gamma = d(\log_{10} I)/d(\log_{10} R)$ beyond 25 kpc of the surface brightness profile as a function of stellar mass and galaxy type. 
At these radii, the surface brightness profiles are not significantly affected by the PSF. The error bars include the variance of the shape
of the surface brightness profile of the galaxies in the stack estimated through bootstrapping.
For low concentration galaxies, the outer slope steepens from $\Gamma \sim -2.5$ at low stellar masses to $\Gamma \sim -4.4$  at higher stellar masses.
For high concentration galaxies, the outer slope steepens from $\Gamma \sim -2.3$ at low stellar masses to $\Gamma \sim -3$  at higher stellar masses.
At fixed mass, the outer slopes of the profiles of low concentration galaxies are steeper than those of high concentration galaxies.
The variance in the slope is much larger for low concentration than high concentration galaxies. Similarly the variance in the slope is much 
larger for low-mass than high-mass galaxies. 

\cite{Ibata} analyze the power-law slope of the two-dimensional projected distribution
of star counts in M31 and find $\Gamma =  -2.30 \pm 0.02$.
We again caution the reader that in stacking large number of galaxies together with different concentrations, 
the resulting  outer slope is a linear combination of the outer slopes 
of the individual galaxies which go into the stack, so our results are not directly comparable to
those obtained for individual galaxies. 
\begin{figure}
  \includegraphics[width=\linewidth]{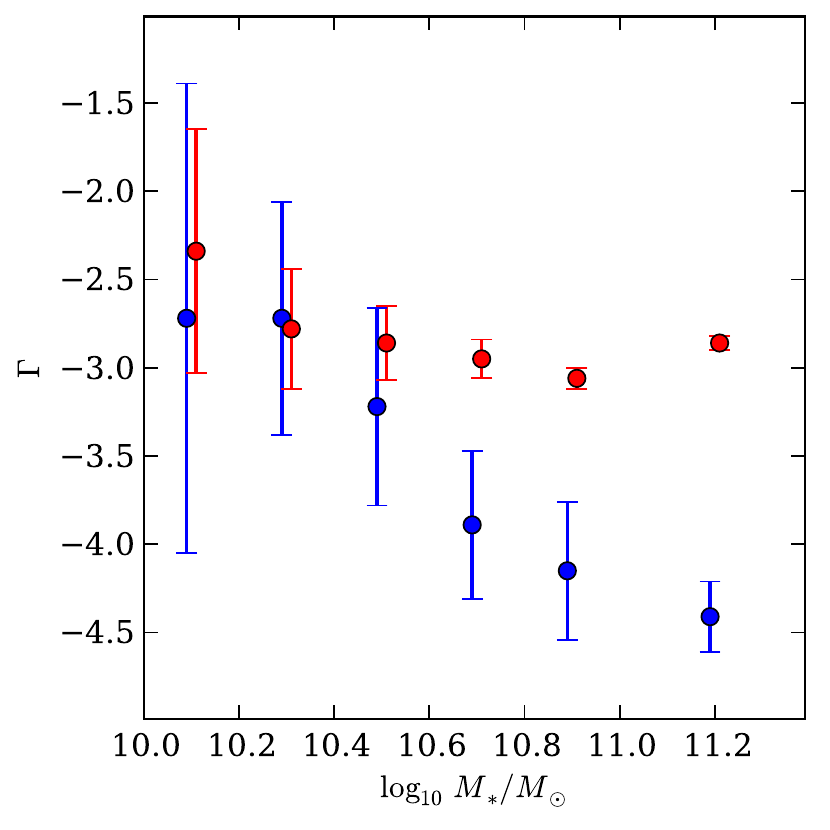}
  \caption{The slope $\Gamma = d(\log_{10} I)/d(\log_{10} R)$ of the surface brightness profile beyond 25 kpc.
    Blue represents low concentration galaxies, while red represents high concentration galaxies. The errors represent
    the total variance in the slope of the surface brightness profile estimated from bootsrapping the samples in the stack.}
  \label{fig:slope_lc_uc}
\end{figure}

\subsection{Colour Profiles as a function of Stellar Mass and  Concentration}
The $g$-$r$ colour profiles extend out to 15-35 kpc for low concentration galaxies and up 
to 40-70 kpc for high concentration galaxies. There also appears to be 
a clear separation between the  inner ($R < 10 \, \mathrm{kpc}$) colour profiles,
where $g$-$r$ decreases as a function of radius, to a region where colour remains
more constant. This is seen for both low and high concentration galaxies. Low concentration
galaxies show steeper inner colour gradients than high concentration galaxies. 
The colour gradient is also steeper in low concentration galaxies with high stellar masses 
than in low concentration galaxies with low masses (See also \citealt{Gonzalez-Perez}, \citealt{Tortora} and \citealt{Suh}). 

For low concentration galaxies, there appears to be a minimum in the $g$-$r$ 
colour beyond which the colour profiles redden. This minimum occurs between 10 kpc for low mass
galaxies and 20 kpc for higher mass systems. For high concentration galaxies, the colour profiles flatten, but do
not exhibit a pronounced upturn. 
This is  consistent with the flattening in colour profiles detected in LRGs \citep{Tal}. 
Reddening of the colour profile at large radii cannot be attributed either 
to the difference in the PSF in the $g$ and $r$ bands or 
due to the errors in the background subtraction. 
The colour profiles of low concentration galaxies do not probe the area where the stellar halo becomes dominant. 
\cite{Bakos} have shown that 90\% of the light profiles of the disks of late-type galaxies exhibit deviations 
from a pure exponential either as truncations (60\%) or as anti-truncations (30\%). 
The colour profiles of disks with truncations are  ``U-shaped''. Disks with anti-truncations exhibit a  plateau
in $g$-$r$ colour at large radii. When stacking a large number
of low concentration galaxies together containing with  a minimum or  a flattening in the $g$-$r$ colour profile, 
the combined effect results in behaviour intermediate between the two. 
Deeper data is required to probe the colours of
stellar populations in  the stellar halo. \cite{Monachesi} detect a flattening of the colour profile of the stellar halo of M81.

The presence of bluer colours in the outer end of both low and high concentration 
galaxies as compared to the centre of the galaxy may 
indicate the presence of stars with significantly younger populations
in these outer parts. However, it will be difficult to confirm this without being able to break 
the degeneracy between age and metallicity by using colours that involve either the $i$ or $z$ bands.

\begin{figure}
  \includegraphics[width=\linewidth]{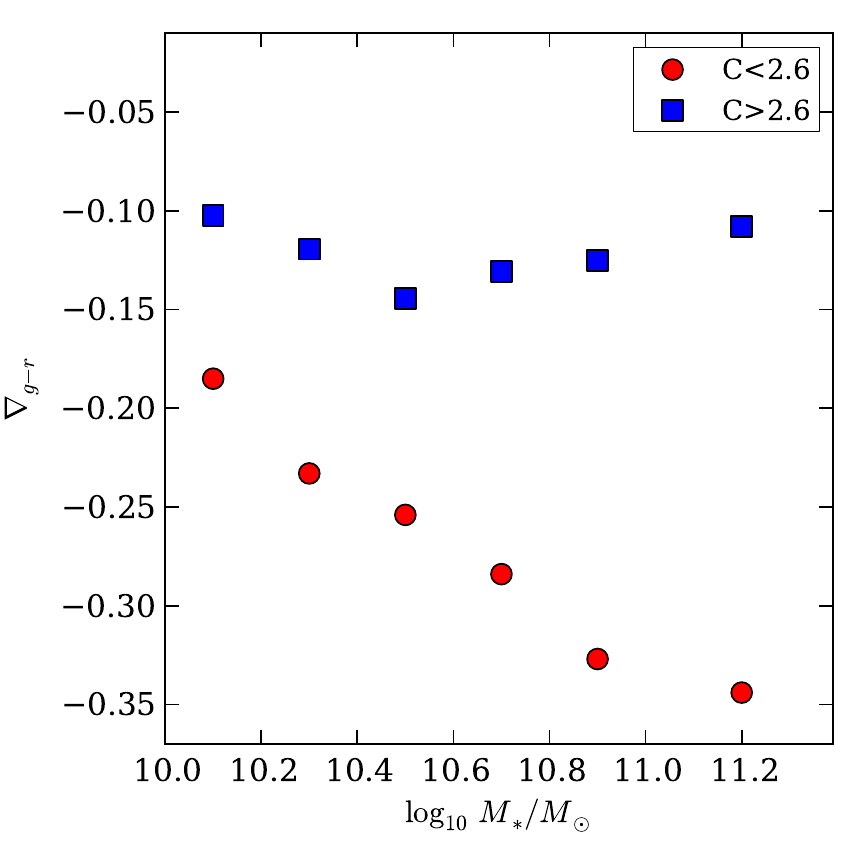}
  \caption{The gradient in the $g$-$r$ colour profile, $\nabla_{g-r} = \frac{\Delta (g-r)}{\Delta (\log_{10} R)}$,  measured along the path of the 
    steepest descent for low concentration galaxies interior to the radius where the
profile exhbits an upturn, and for high concentration galaxies interior radius where the colour profile flattens.}
  \label{fig:colour_gradient}
\end{figure}

We plot the $g$-$r$ colour gradient  $\nabla_{g-r} = \frac{\Delta (g-r)}{\Delta (\log R)}$ for our 
galaxy stacks in Figure \ref{fig:colour_gradient}. For low 
concentration galaxies, we evaluate the slope  for the path of the steepest descent interior 
to the minimum in the $g$-$r$ colour profile. For high concentration galaxies, the slope is      
derived for the steepest descent interior to the point where the $g$-$r$ colour profile flattens. 
Since the colour profile is affected by the PSF at the centre
of the galaxy stack, the analysis is restricted to radii beyond 3 kpc.  
The gradient is first evaluated from 3 kpc right up to the minimum in the $g$-$r$
colour profile, and the path length over which the gradient is calculated
is decreased step-by-step until the gradient reaches its maximum.
Figure \ref{fig:colour_gradient} shows that  colour gradients are stronger in late-type galaxies
than in early-type galaxies. In early-type galaxies, the gradients do not depend on stellar mass,
but in late-type galaxies, high mass galaxies have much steeper gradients than low mass galaxies.  

\section{Multi-Component Modelling of the Galaxy}
\label{sec:modelling}
For each stacked image, we  model the full two dimensional $r$-band intensity distribution of the 
galaxy stack using multi-component S\'{e}rsic models. 
We are particularly interested in  modelling the outer stellar halo light of the galaxy and in placing constraints on the
amount of accreted stellar material. We are confident that the depth of our 
stacked images means that we can reach out into the extended stellar 
halo of the galaxy. Theoretical considerations indicate that there should be an inflexion or 
a change in the surface brightness profile of the galaxy where the accreted stellar component begins to dominate \citep{Cooper}.

The \cite{Sersic} profile $\log I(r) \propto r^{1/n}$ is the most versatile among the 
models and is traditionally used to fit the surface brightness profile of galaxies.  
The S\'{e}rsic profile reduces to an exponential ($n=1$) profile for disk galaxies, 
while $n=4$ profiles \citep{deVauc} has been used to model bulges and ellipticals. 
\cite{Kormendy} have demonstrated that the S\'{e}rsic profile fits elliptical galaxies and spheroidals 
very well over a large dynamic range in radius. They also suggested that departures 
from these profiles could provide new insights into galaxy formation. In this paper, we leave aside the issue of  
departures from the S\'{e}rsic profile at small radii in our galaxy stacks.
Our aim is to explore our ansatz that the excess light (deviations from the single 
S\'{e}rsic profile) detected at large radii ($R>20\,\mathrm{kpc}$) 
is indicative of additional components in the galaxy, which may be attributed to  
accreted stellar material. Our second ansatz is that the radial variation of ellipticity 
can also be indicative of various galaxy formation processes. In particular, the difference in 
ellipticity between the inner part of the galaxy and the outer stellar halo of the galaxy
may yield clues to the origin of these components. 

Deviations from simple profiles at large radii and the radial variation in ellipticity can be adequately modelled through multi-component modelling, 
where each component can be represented by a S\'{e}rsic profile with a fixed ellipticity. The flexibility of the S\'{e}rsic profile helps us 
model a large variety of possible profiles. The real challenge of modelling galaxies is in assigning a physical significance to each of these components. In fitting 
multiple components to our galaxy stacks, we are motivated by the results of \cite{Cooper} who have demonstrated theoretically from particle-tagging methods that the 
in-situ and the accreted surface density profiles are well fit by \cite{Sersic} functions, while the total profile is best fit by a sum of these two functions.

We seek to model the two-dimensional intensity profile of the galaxy with a minimum number of components. In the following subsections, we model separately the 
stacks of high concentration and low concentration galaxies. We first show that a single component is not sufficient to model the surface brightness profile of high 
concentration galaxies. We demonstrate how the surface brightness profile of high concentration galaxies can be successfully modelled by two components. For low concentration 
galaxies, we show that we may need three components to model the disk breaks of galaxies in addition to the stellar halo. For all our fitting procedures, we use the 
full two-dimensional information in the stacked image. We also test our modelling on mock images of high and low concentration galaxies.

\subsection{High Concentration Galaxies}
High concentration galaxies are  simpler to model than low concentration galaxies. Motivated by this, we first fit a 
single two-dimensional S\'{e}rsic model with a fixed ellipticity to our high concentration galaxy stack:
\begin{equation}
I(R)= I_{0} \exp \left\{ -b_{n} \big( \left( \frac{R(q)}{R_{e}} \right) ^{1/n}  - 1 \big) \right\},
\end{equation}
where $I_{e}$ is the intensity at the effective radius $R_e$ that encloses half of the total light from the model and $n$ is the S\'{e}rsic index.
The constant $b_{n}$ is defined in terms of the S\'{e}rsic index. The radial distance, $R$, is a function of the Cartesian coordinates and the 
ellipticity $q$ of the model. We also model an additional constant sky component. A single S\'{e}rsic model so defined has a total of 4 free parameters. 

We compare this with a double S\'{e}rsic model with a common centre and with different ellipticities for each S\'{e}rsic component.
S\'{e}rsic profiles extend out to infinity. In order to ensure that the outer stellar halo is determined by only one component, we smoothly 
cut off the inner S\'{e}rsic profile at large radii: the surface brightness profile is suppressed beyond $7\,R_{eff}$ and drops 
to zero outside $8\,R_{eff}$.\footnote{The same procedure is followed in SDSS for pure de Vaucouleurs profile to calculate \texttt{ModelMag}.} 
With an additional constant sky component ($c$), the double S\'{e}rsic model has a total of 9 free parameters.
There are two additional free parameters for the centre of each model. To reduce the number of free parameters, we determine and fix 
the centre of the galaxy stack by fitting a single S\'{e}rsic model with variable parameters for the centre. All the models considered are 
symmetric along the major axis and the minor axis. The asymmetries in the image (in the form of bars, bulges, disks, pseudo-bulges, etc.)
are not explicitly modelled and appear as residuals.

For the fitting procedure, each model was convolved with the average stacked SDSS PSF before fitting (see section \ref{section:psf}). 
We employ a Bayesian technique with uniform and physical priors for all the parameters $\theta$ ($I_{0}:\,0$--$1\mathrm{\,\,nanomaggies\,arcsec}^{-2};\,R_{e}:\,1$--$100\mathrm{\,\,pixels;}\, 
n:\,0$--$10;\, c:\,0$--$1\mathrm{\,\,nanomaggies\,arcsec}^{-2}$;$\,q:\,0-10$). 

Applying Bayes' theorem, we can find the posterior probability distribution over the parameters $\theta$ as 
\begin{equation}
p(\mathbf{\theta} \mid \mathbf{D}) = \frac{p(\mathbf{D} \mid \mathbf{\theta})}{\int_\mathbf{\theta} p(\mathbf{D} \mid \mathbf{\theta}) p(\mathbf{\theta}) \, d\mathbf{\theta}} \cdot p(\mathbf{\theta}), 
\label{equation:bayes}
\end{equation}
where  $\int_\mathbf{\theta} p(\mathbf{D} \mid \mathbf{\theta}) p(\mathbf{\theta}) \, d\mathbf{\theta}$ is the model evidence and $\mathbf{D}$ is the data. 

$p(\mathbf{D} \mid \mathbf{\theta})$ is the likelihood which can be constructed as follows:
\begin{equation}
\log(L)= -\frac{1}{2} \log ((2\pi)^k\Sigma\,) -\frac{1}{2}(\mathbf{D}-\mu(\theta))^{\rm T}\Sigma^{-1}(\mathbf{D}-\mu(\theta)),
\label{equation:likelihood1}
\end{equation}
where $\Sigma$ is the covariance matrix (which is diagonal in this case), $\mathbf{D}$ is the stacked data, 
$\mu$ is the model as a function of the parameters $\theta$ and $k$ is the number of independent pixels.

We use {\sc Multinest} \citep{Multinest1,Multinest2}, a Bayesian inference tool on the full stacked image. This has the advantage over Galfit \citep{Galfit} 
in that it can explore the complete parameter space. We use the full image $950 \times 950$ pixels for the fitting procedure. This is essential for a 
proper determination of the residual sky component in the stacked images. In general, the determination of the outer S\'{e}rsic index is correlated with 
the sky component. 

We generate a full posterior probability distribution function (PDF) of all the parameters using {\sc Multinest}. This allows us to evaluate
the degeneracies in the parameters. If the posterior PDF is double modal (i.e., contains two maxima), we choose the most physical
model such that the effective intensity ($I_{e}$) / effective radius ($R_{e}$) of the inner most component should be larger/smaller
than that of the outer component. For the final parameters of the model, we use the mean values of the posterior PDF. These mean
values automatically encode information on the parameter degeneracies.

To compare the various models with each other, we can use two approaches. The first involves using the Bayesian ``evidence'' 
marginalised over the model parameters for model comparison. This compares models on a global scale. On the other hand, comparing 
residuals (or the reduced chi-square) in specific regions of the stacked image allows one to judge the goodness of fit 
for specific components of the galaxy stack including the stellar halo.

To compare models globally, we construct the Bayes factor ($B_{10}$ - hypothesis 1 over hypothesis 0). Kass and Raferty (1995, Journal of 
American Statistical Association) suggest comparing $2\log_{e}(B_{10})$ and note that a factor $>10$ is indicative of strong evidence against hypothesis 0.
The square root of $2\log_{e}(B_{10})$ gives us the level of significance between the two models. We compare the factor $2\log_{e}(B_{D/S})$ which is comparing
the double S\'{e}rsic model over the single S\'{e}rsic model for a range of mass bins 
in Table \ref{table3}. In Figure \ref{fig:lum_compare_models}, we show      
how well the double S\'{e}rsic model fits the surface brightness profiles for a range of stellar mass bins.

\begin{table}
\caption{We compare the double-S\'{e}rsic model with the single-S\'{e}rsic model by comparing $2\log_{e}(B_{D/S})$, where $B_{D/S}$ is the Bayes factor favouring
the double-S\'{e}rsic model over the single-S\'{e}rsic model}
\label{table3}
\begin{tabular}{ c | c}
Mass bin & $2\log_{e}(B_{D/S})$\\
\hline
10.0-10.2 & 3713  \\
10.2-10.4 & 9779 \\
10.4-10.6 & 23508 \\
10.6-10.8 & 30831 \\
10.8-11.0 & 21730 \\
11.0-11.4 & 18727 \\
\hline
\end{tabular}
\end{table}

\begin{figure}
\includegraphics[width=\linewidth]{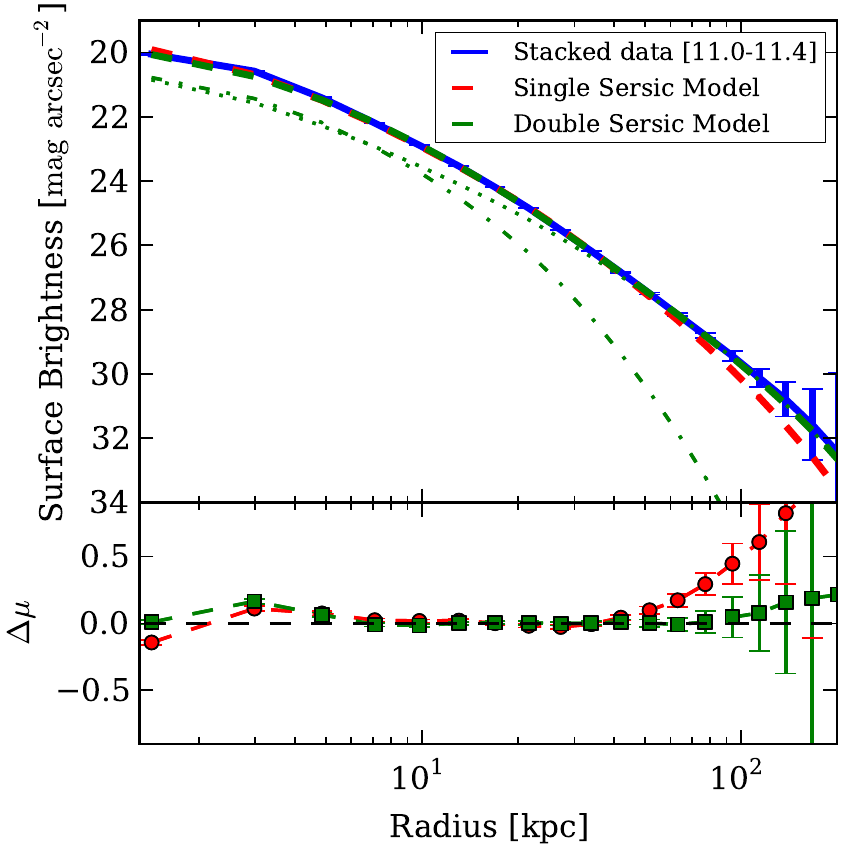}
\caption{Comparison of the the double-S\'{e}rsic (dashed green) and the single-S\'{e}rsic 
(dashed red) models with the surface brightness profile of the high concentration highest stellar mass bin stack 
$C>2.6$,$10^{11.0} \msun < M_{*} < 10^{11.4} \msun$). For the double-S\'{e}rsic model, 
the internal component is denoted by the dot-dashed line while the outer component is denoted by the dotted line.}
  \label{fig:model_fits}
\end{figure}

In Figure \ref{fig:model_fits}, we compare the single-S\'{e}rsic and double-S\'{e}rsic models for the high concentration galaxy 
stack in the highest stellar mass bin $10^{11.0} \msun < M_{*} < 10^{11.4} \msun$. The single-S\'{e}rsic function fits the symmetric 
central high S/N part of the surface brightness profile up to a surface brightness of $\mu_r \sim 27 \,\mathrm{mag\,arcsec}^{-2}$  
reaching out to 30 kpc. Note that all  internal galaxy components (e.g. bulges, disks, pseudo-bulges) are averaged out and incorporated 
into the single-S\'{e}rsic fit. Beyond 30 kpc, excess light is detected. 
The double-S\'{e}rsic profile on the other hand provides an excellent fit 
up to a depth of $\mu_r \sim 32 \,\mathrm{mag\,arcsec}^{-2}$ reaching out to 130 kpc. The residuals are shown in the panel below 
in Figure \ref{fig:model_fits}. The residuals of the double-S\'{e}rsic are less than $0.2 \,\mathrm{mag\,arcsec}^{-2}$ across 
the whole radial range (0-120 kpc) of the galaxy stack. The residuals at the centre are attributed to the asymmetric 
part of the intensity distribution at centre of the galaxy stack due to the various internal galaxy components mentioned 
above. 

\begin{figure}
  \includegraphics[width=\linewidth]{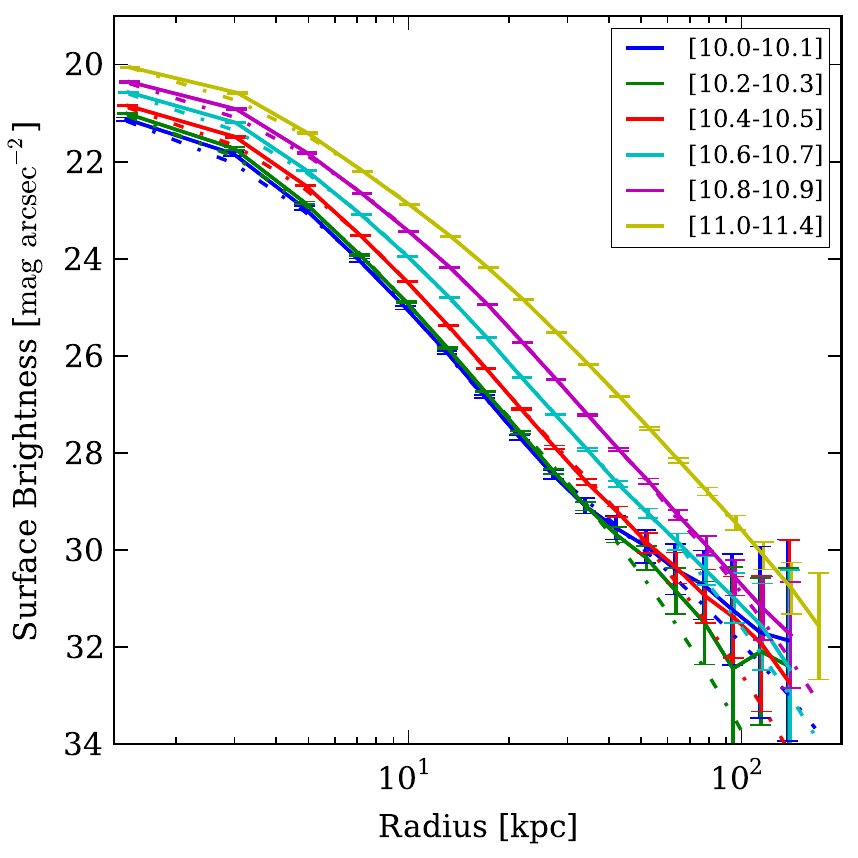}
  \caption{We compare the surface brightness profiles of the various stellar mass bins with the double-S\'{e}rsic models.}
  \label{fig:lum_compare_models}
\end{figure}

We find that the double-S\'{e}rsic profile provides a much better fit for all high concentration galaxies across all mass bin ranges.
This can be seen visually by calculating and comparing the residuals of the image beyond $20$ kpc for each model.
Significant deviations are only seen in the lower two mass bins. 
The fits to  the lowest mass bin is not perfect due to limited number of galaxy images ($ \sim 1212$) which went into the stack.
At first glance, our conclusion that a double-S\'{e}rsic profile is {\em always}
required  may seem surprising, because the surface brightness profiles of massive galaxies
with high concentration do not exhibit a clear inflexion point.  
We note that a single-S\'{e}rsic model has a single fixed ellipticity, while the double-S\'{e}rsic model with different ellipticities 
for each component can in a limited way mimic the varying ellipticity of the stacked galaxy image. 
We  investigated  whether the change in ellipticity is the dominant factor that favours a double-S\'{e}rsic profile over a 
single-S\'{e}rsic profile. To test this, we compare a single-S\'{e}rsic and a double-S\'{e}rsic profile fitted to similar stacks of galaxies 
which are not aligned but are randomly oriented. In all cases, the double-S\'{e}rsic is still preferred over the single-S\'{e}rsic profile. 
The factor $2\log_{e}(B_{D/S})$ in the randomly oriented case is reduced to one-third of that as calculated in Table \ref{table3}. This indicates 
that it is both the surface brightness profile and the ellipticity which contribute to favour a double-S\'{e}rsic profile over a single-S\'{e}rsic profile.

\begin{figure}
  \includegraphics[width=\linewidth]{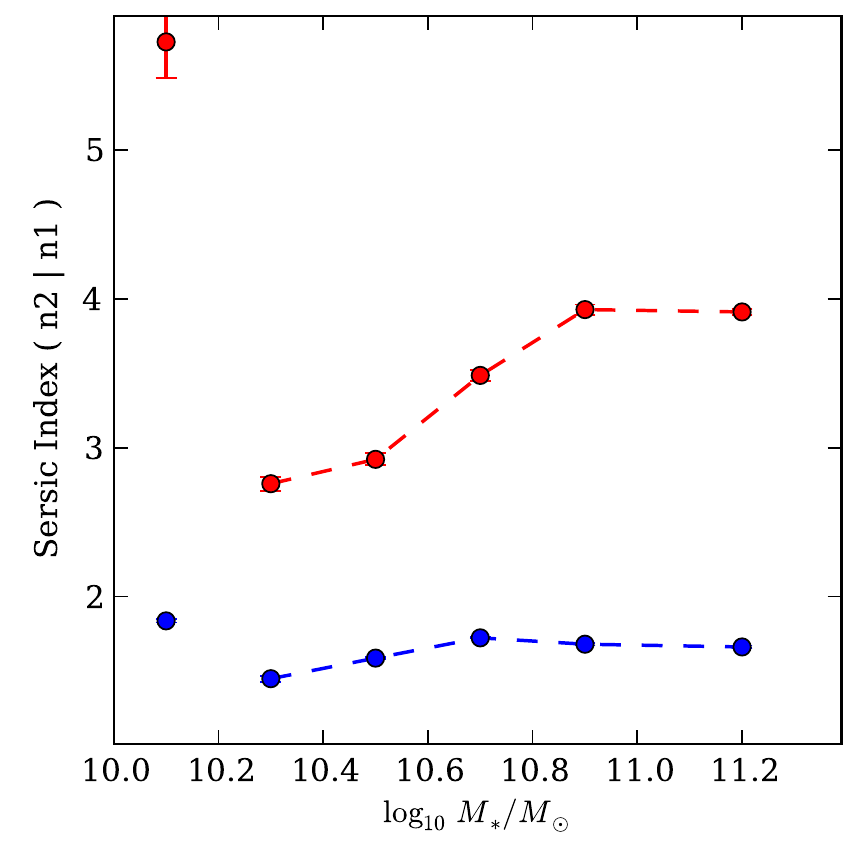}
  \includegraphics[width=\linewidth]{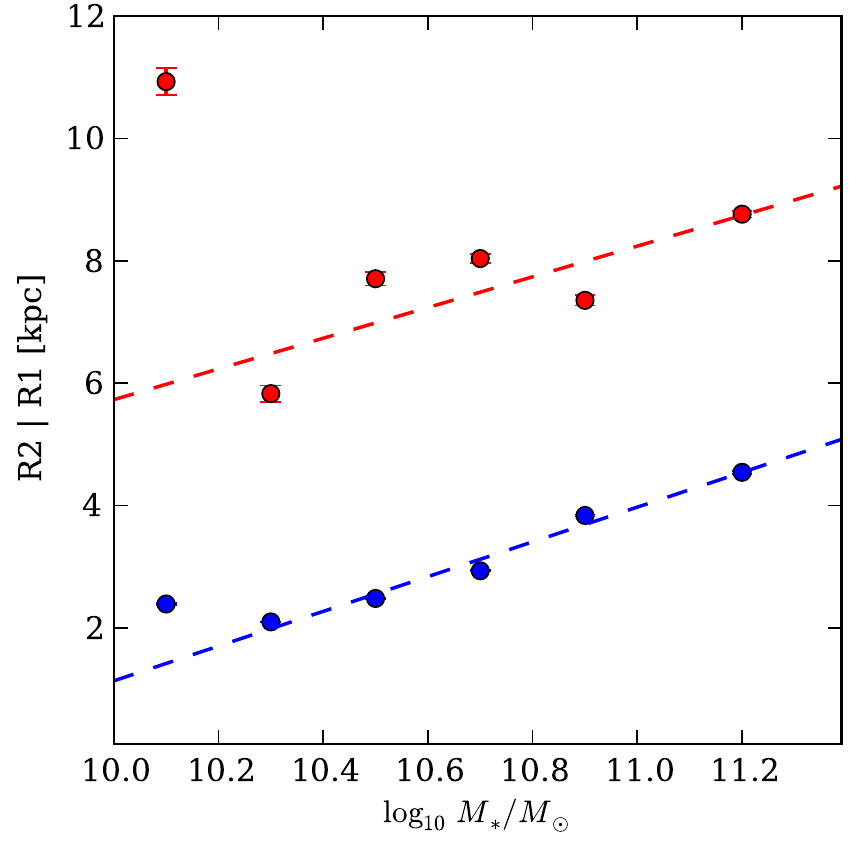}
  \caption{(a) The S\'{e}rsic indices of the inner(blue) and outer(red) components for high concentration galaxies. (b) The effective radii of the inner(blue) 
    and outer(red) S\'{e}rsic components for high concentration galaxies. The outer effective radius scales as $\propto 2.5 \log_{10} M_{*}$ while the inner effective
    radius scales as $\propto 2.8 \log_{10} M_{*}$. The model fails to fit for the lowest mass bin because of insufficient numbers in the stack.}
  \label{fig:conc_n1n2}
\end{figure}

Plots of the  S\'{e}rsic indices of the two components
as a function of mass are shown in in Figure \ref{fig:conc_n1n2}. The outer S\'{e}rsic index increases with
the mass of the galaxy stack from $n \sim 3$ to $n \sim 4$.  
The effective radii of each component are also denoted 
in the Figure \ref{fig:conc_n1n2}. The effective radius of the outer 
component scales as $\propto 2.5 \log_{10} M_{*}$ reaching a maximum 
of 9 kpc for the highest mass bins. We note that the  inner S\'{e}rsic 
component is always  more elliptical than the outer S\'{e}rsic component. 
The ellipticity of the inner component is approximately constant for all mass bins while the ellipticity of the outer component increases 
as mass increases.

\begin{figure}
  \includegraphics[width=\linewidth]{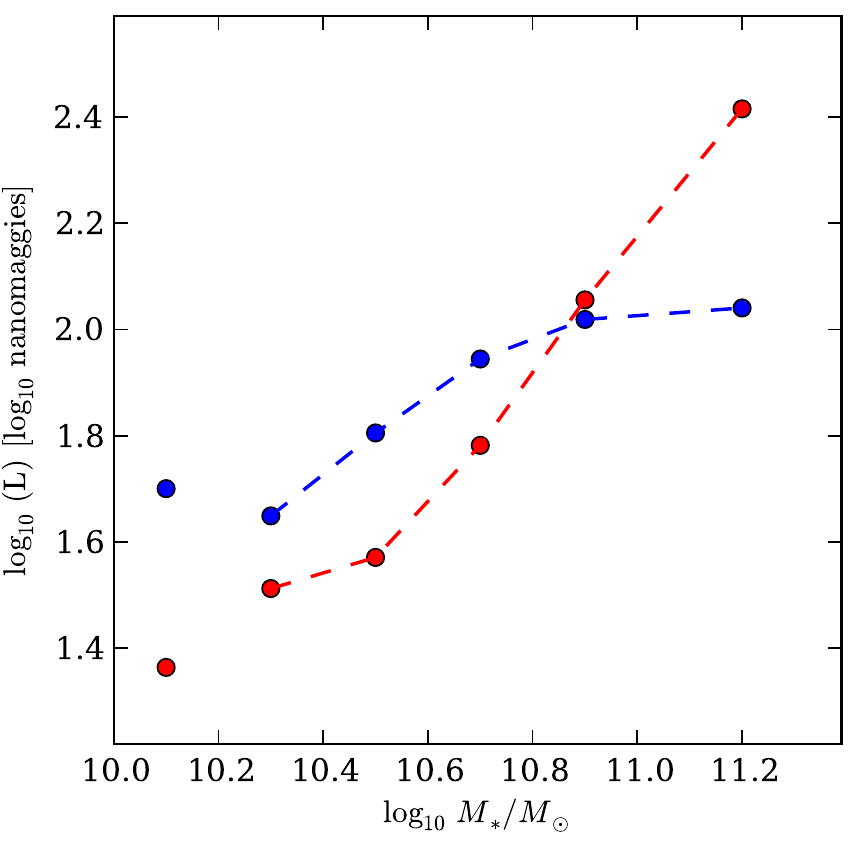}
  \includegraphics[width=\linewidth]{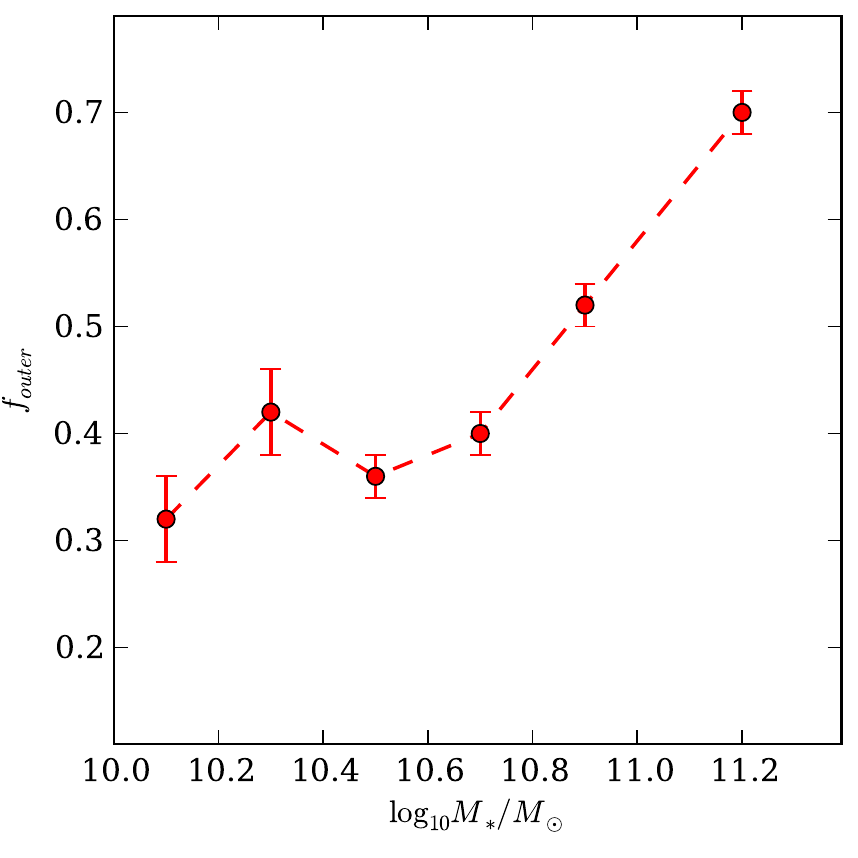}
  \caption{(a) The $\log_{10}$ of the $r$-band total Luminosity (in nanomaggies) in the inner(blue) and outer(red) components as a function of stellar mass for high concentration galaxies. 
    (b) The fraction of light in the outer S\'{e}rsic component as function of stellar mass for high concentration galaxies.}
  \label{fig:light-fraction_uc}
\end{figure}

Having separated the light from the galaxy into two components, we study the variation of the light in the two components as a function of 
stellar mass. We can also calculate the fraction of light in the 
outer S\'{e}rsic component (Figure \ref{fig:light-fraction_uc}). We will discuss this result in Section \ref{sec:discussion}.

\subsection{Low Concentration Galaxies}
Modelling low concentration galaxies along with their stellar halo component remains a challenging task,  because of the 
extremely low fraction of light in the stellar halo in these systems. Estimates of the stellar halo contribution  for M31 lie 
between $0.6$ and $1.5$ percent \citep{Ibata}, while those for the Milky Way lie 
between $0.3$ and $1.0$ percent \citep{Bell,McMillan}.
Previous modelling and estimates of the stellar halo content of disk galaxies have been made from star counts. 
In order to detect the stellar 
halo in face-on disk galaxies, deep imaging is necessary with an accurate determination of the 
background residuals. Recently \cite{dokkum}
tried to model and determine the stellar halo content of the massive spiral galaxy 
M101 from integrated surface brightness profiles by going
to a depth of  $\mu_g \sim 32 \,\mathrm{mag\,arcsec}^{-2}$.  
The effective depths of our stacked images are similar to this.

\begin{figure}
  \includegraphics[width=\linewidth]{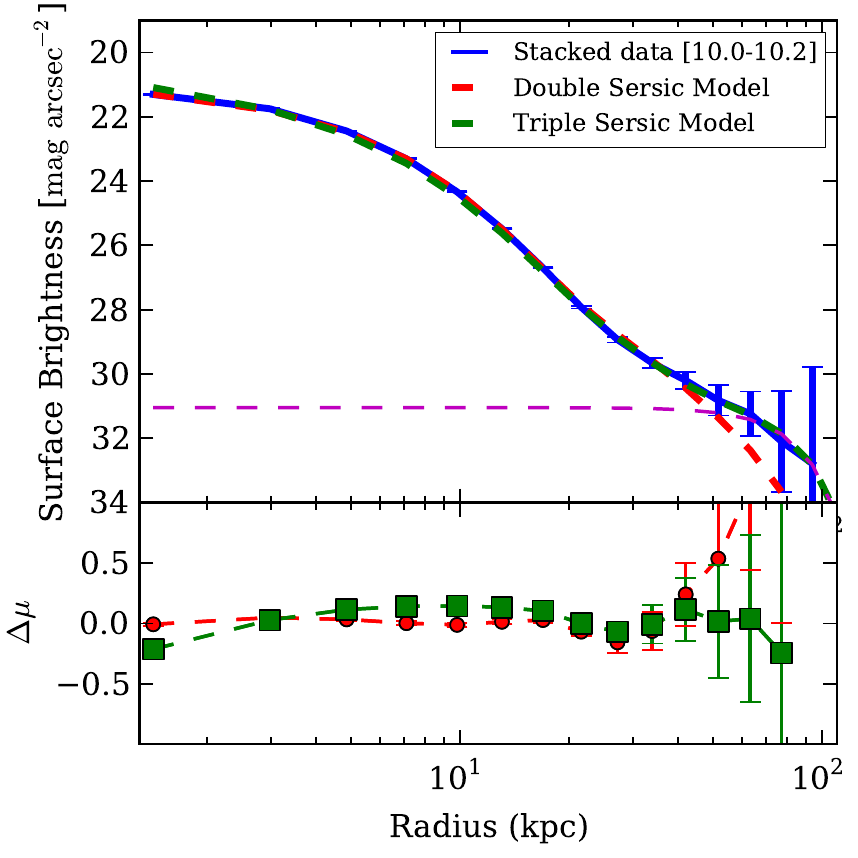}
  \caption{The double S\'{e}rsic model (shown in red) provides an inadequate fit to low concentration ($C<2.6$) low mass galaxies ($10^{10.0} \msun < M_{*} < 10^{10.2} \msun$)
    The triple S\'{e}rsic model provides a much better fit (shown in green). The third component of the triple S\'{e}rsic model is shown in a dashed magenta line. In the bottom
    panel, the residuals of the double S\'{e}rsic model and the triple S\'{e}rsic model are shown.
  }
  \label{fig:diskgal_doublesersic}
\end{figure}

Another important issue is that  disk breaks in galaxies \citep{Bakos} also cause inflections 
in the surface brightness profile of the stacked 
galaxies and need to be modelled. We find that a double-S\'{e}rsic model often fails to fit the stacks 
of low concentration galaxies,
as is shown in Figure \ref{fig:diskgal_doublesersic}. In these galaxies, the inflection is caused by disk breaks 
and these breaks can occur very close to where the stellar halo becomes dominant.

A natural extension of our modelling procedure would be to use a concentric triple S\'{e}rsic model. 
However, the general triple S\'{e}rsic model is 
highly degenerate, especially when trying to separate components which are not easily distinguishable from each 
other. Face-on disk galaxies with a low stellar-halo mass fraction occupy 
only a limited parameter space of a three component model. To break these degeneracies, we truncate    
the inner two components (beyond $7-8\,R_{e}$) and apply 
restrictions to the third component of the triple S\'{e}rsic model. In particular, we look for 3rd component  
solutions that involve a low S\'{e}rsic index ($n3\,<\,1.5$), lower effective intensity (in comparison 
to the other 2 components) and a larger effective radius ($R_{eff}> 15\,\mathrm{kpc}$) for the outer-most component. 
The low S\'{e}rsic index ensures that the profile of the 
third outer component does not rise steeply and dominate the inner central parts of the galaxy.

We also modify our fitting algorithm as follows. We do not fit three components at the same time. 
We first model independently the galactic disk along with the disk break in
high S/N part of the stacked image with a truncated double S\'{e}rsic model. Later, having fixed the two components
describing the internal part of the galaxy, we add a  
third component to model the outer extra light. This is necessary because the S/N of the light of the
outer image is so much lower than that of the inner regions.  
If the disk break occurs close to the where the stellar halo becomes dominant 
(i.e., if the stellar halo fraction is not negligible), we first model the 
internal two components with a truncated double 
S\'{e}rsic model. Then keeping the innermost component fixed, 
we model the disk break and the extra stellar halo light 
by fitting two additional S\'{e}rsic components. In both methods, we determine the constant sky component at each step. 

\begin{figure}
  \includegraphics[width=\linewidth]{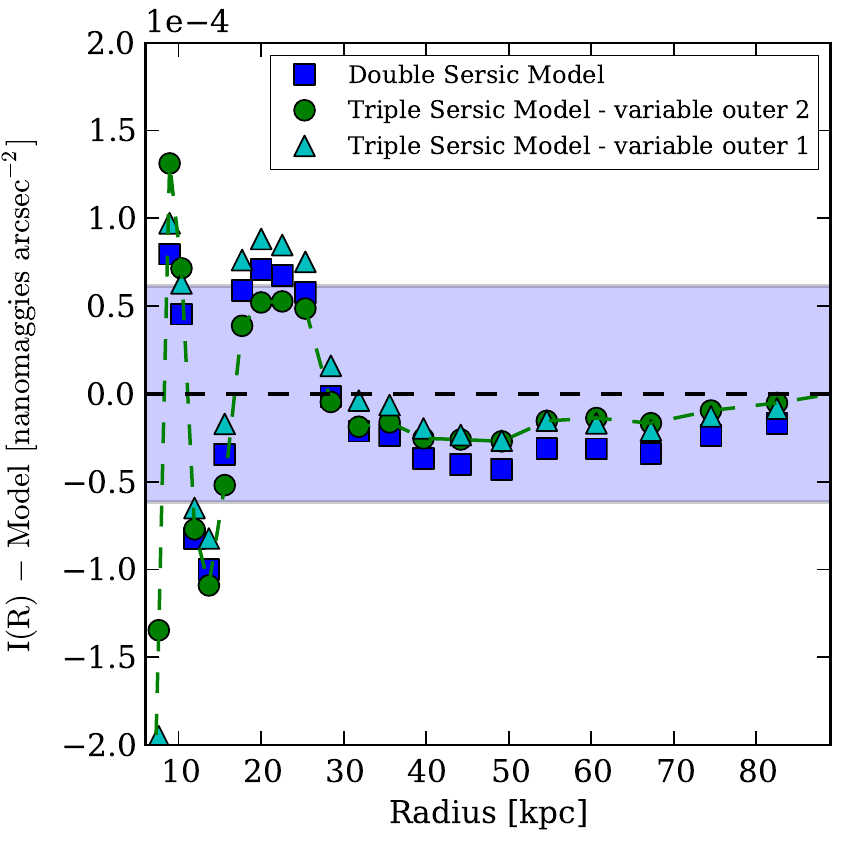}
  \caption{The residuals for three models: the double S\'{e}rsic, the triple S\'{e}rsic by keeping the inner most component fixed (Method A) and the triple S\'{e}rsic
    by keeping the inner two components fixed (Method B) for galaxies stacked in the mass bin range $10^{10.0} \msun < M_{*} < 10^{10.2} \msun$ and with concentration $C<2.6$.
    The fraction of light of the galaxy in the outermost component by Method A is $2.3\pm 0.4\%$ and by method B is $1.2 \pm 0.3\%$. The blue band gives the average uncertainty
    in the background removal for each pixel in $\mathrm{nanomaggies\,arcsec}^{-2}$.
  }
  \label{fig:compare_models}
\end{figure}

The global Bayes factor is unable to differentiate between models in the low concentration case, since it is dominated
by the asymmetric component (bars, pseudo-bulges, etc.) at the centre of the stacked galaxy. In order to judge which fitting method is most appropriate
for a given  given galaxy stack, we subject every image stack to both methods and calculate the chi-square of the image for 
each pixel beyond $20$ kpc. We compare the reduced chi-square for a double S\'{e}rsic model, as well as 
both methods for determining the third component of the
triple S\'{e}rsic model, and choose the best fit model. 
In Figure \ref{fig:compare_models},  we compare the residuals of 
the double S\'{e}rsic model as well as the two methods for determining the 
components of the triple S\'{e}rsic model for disk galaxies stacked in the mass bin range 
$10^{10.2} \msun < M_{*} < 10^{10.4} \msun$,  with concentration index $C<2.6$. 
The blue band gives the average uncertainty in background removal for each pixel in 
$\mathrm{nanomaggies\,arcsec}^{-2}$. The procedure which keeps 
the the inner most component fixed and varies the outer two components fares the best.
The best fit triple S\'{e}rsic model is shown in Figure \ref{fig:diskgal_doublesersic}.

The accuracy of modelling the third component depends upon the accuracy of the correct background sky determination.
This accuracy is limited by the accuracy of our background removal. 
For the model fits to the stack of $N \sim 3000$ galaxy images shown in Figure
\ref{fig:diskgal_doublesersic},
if we assume a conservative S\'{e}rsic index ($n \sim 0.4$) 
and an effective radius $R_e \sim 40\,\mathrm{kpc}$ and 
an effective magnitude determined by the error of the background residuals 
($I_{e} \sim 6 \times 10^{-5}\,\mathrm{nanomaggies\,arcsec}^{-2}$), the third component can 
be correctly determined if it is greater than $2\%$ of the total light in the galaxy. 

\begin{figure}
\includegraphics[width=\linewidth]{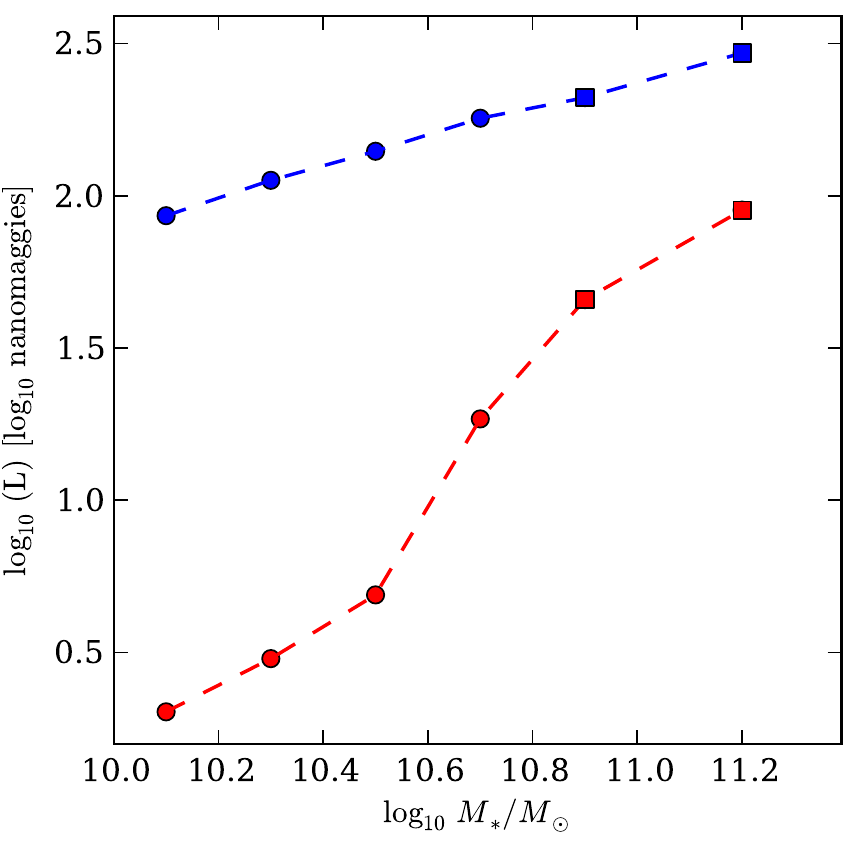}
\includegraphics[width=\linewidth]{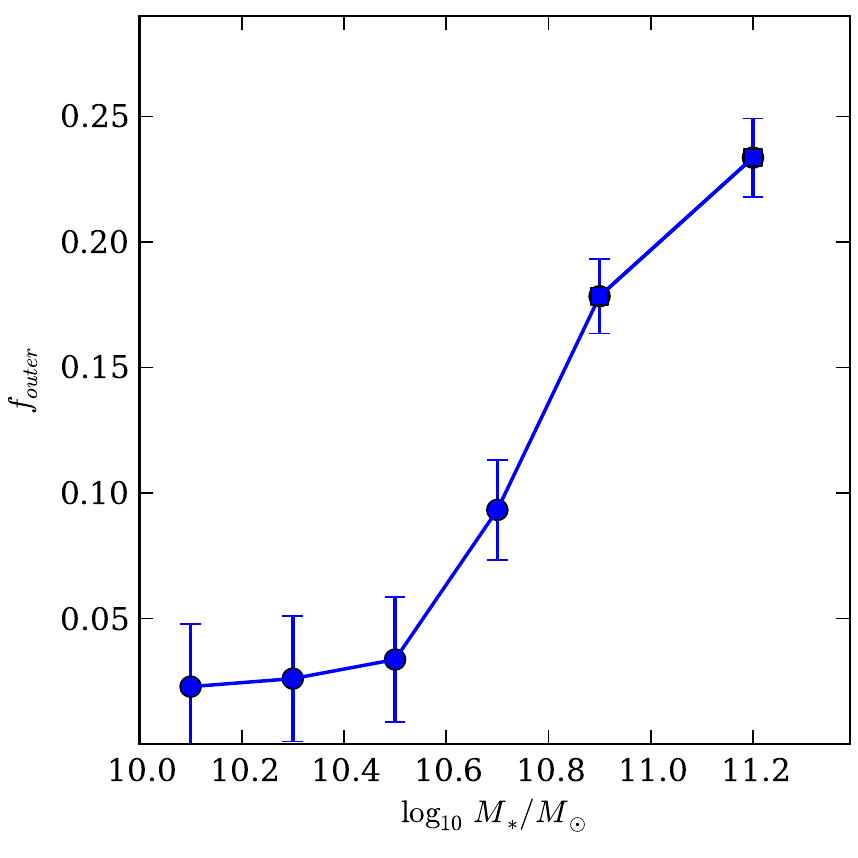}
\caption{(a) The $\log_{10}$ of the $r$-band total luminosity (in nanomaggies) in the inner 
and outer-most components as a function of stellar mass for low concentration galaxies. 
(b) The fraction of light in the outer-most S\'{e}rsic component as a function of stellar mass for low concentration galaxies. 
Circular markers indicate that a triple S\'{e}rsic profile was required to model the outer parts of the stellar halo, while the square markers 
indicate that a double S\'{e}rsic profile was sufficient.}
\label{fig:light-fraction_lc}
\end{figure}

In Figure \ref{fig:light-fraction_lc}, we plot the fraction of the total
light and stellar mass of the galaxy  in the inner and outermost components. 
Results are shown as a function of $M_*$ and for low and high concentration systems.
For low concentration galaxies, the higher two mass bins are best fit by double 
S\'{e}rsic models, while the lower mass bins are best 
fit by  triple S\'{e}rsic models. Most of the low concentration stacks 
which are modelled successfully by a triple-S\'{e}rsic profile are
best fit by keeping only the inner-most component fixed. Only one low 
concentration stack ($10^{10.0} \msun < M_{*} < 10^{10.2} \msun$) could be
best fit by fixing the inner two S\'{e}rsic components. 
We will discuss these results later in Section \ref{sec:discussion}.

\begin{figure}
  \includegraphics[width=\linewidth]{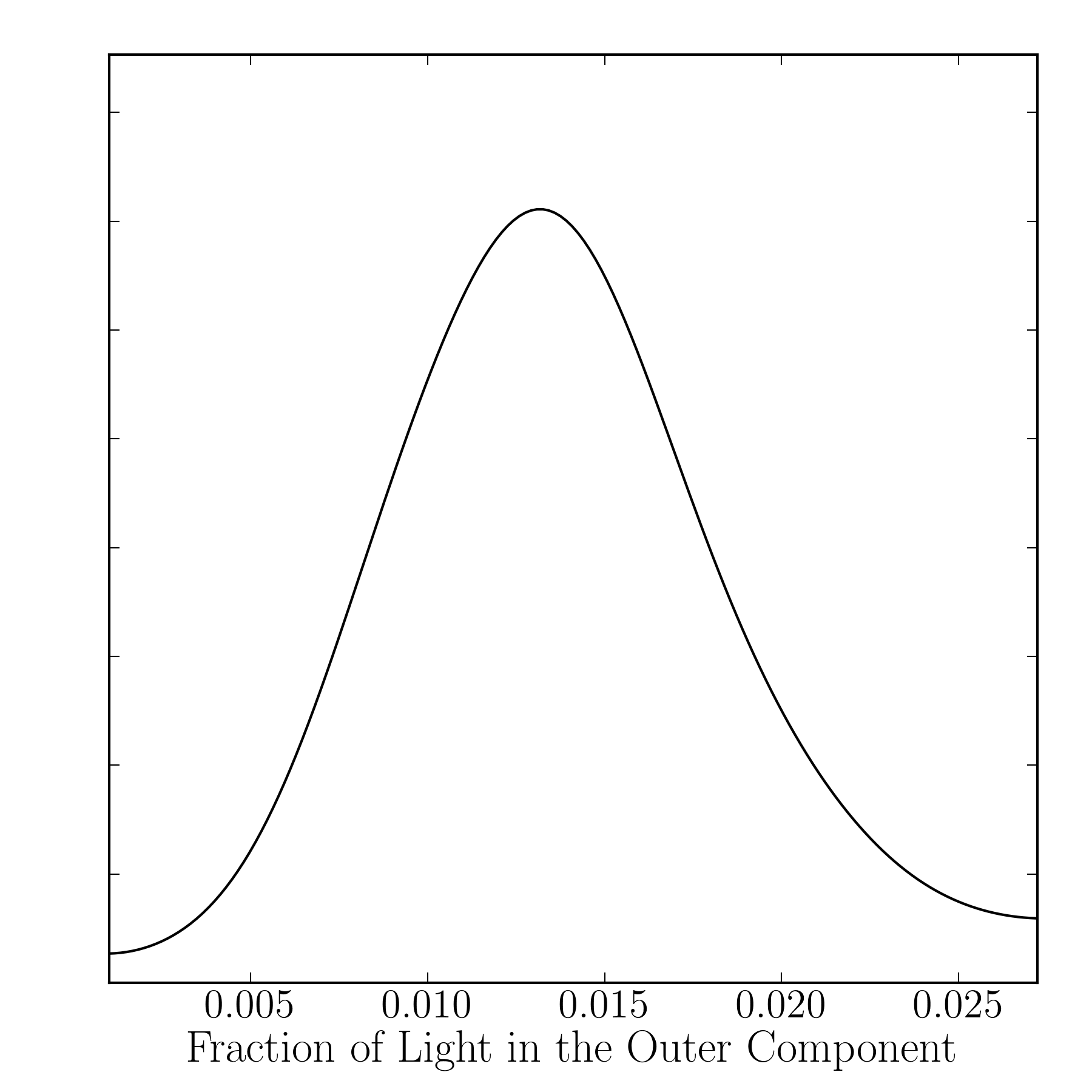}
  \caption{The probability distribution function (PDF) of $f_{outer}$ for the stacked image of disk galaxies in the mass
    range  $10^{10.0} \msun < M_{*} < 10^{10.8} \msun$ and with a concentration of $C < 2.6$.}
  \label{fig:disk_massfraction}
\end{figure}

Improved accuracy in determining the third component may be obtained  by stacking a larger 
number of low concentration galaxies.
We stack 12,423 galaxies in the $r$-band with random orientations in the mass r
ange $10^{10.0} \msun < M_{*} < 10^{10.8} \msun$, 
with a concentration $C < 2.4$ and with an isophotal axial ratio $>=0.77$. 
Using our modelling procedure, we can derive the probability distribution function (PDF) of the 
fraction of light in the third component (see Figure \ref{fig:disk_massfraction}). This fraction is about $1.3 \pm 0.5 \%$.

\begin{figure}
  \includegraphics[width=\linewidth]{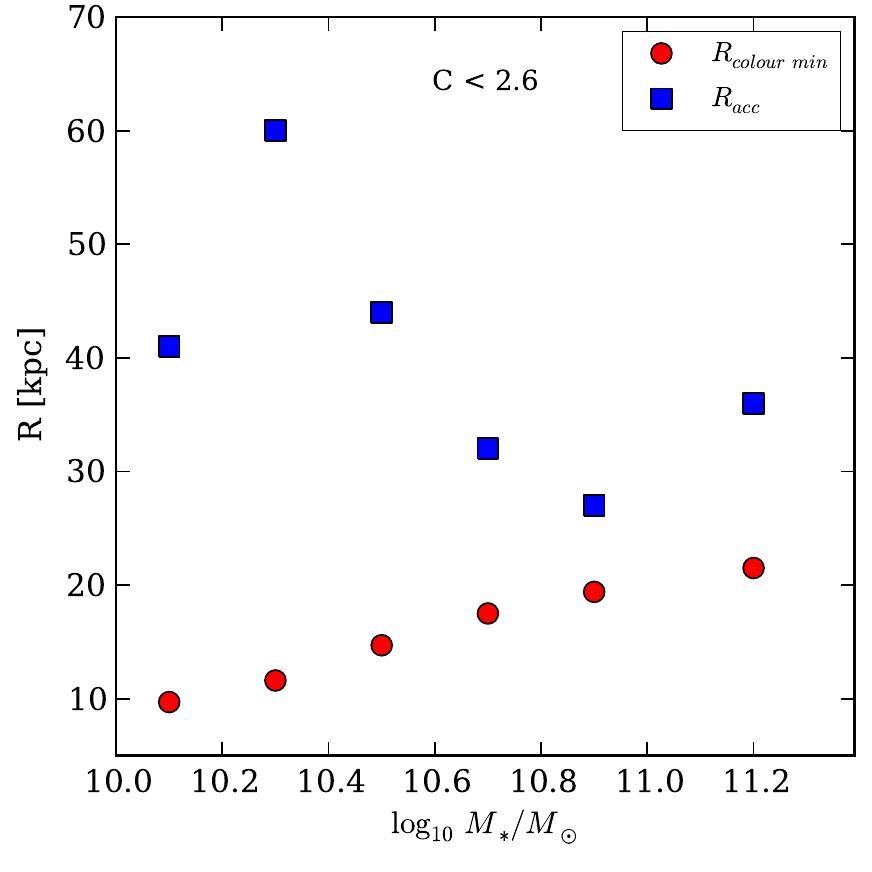}
  \includegraphics[width=\linewidth]{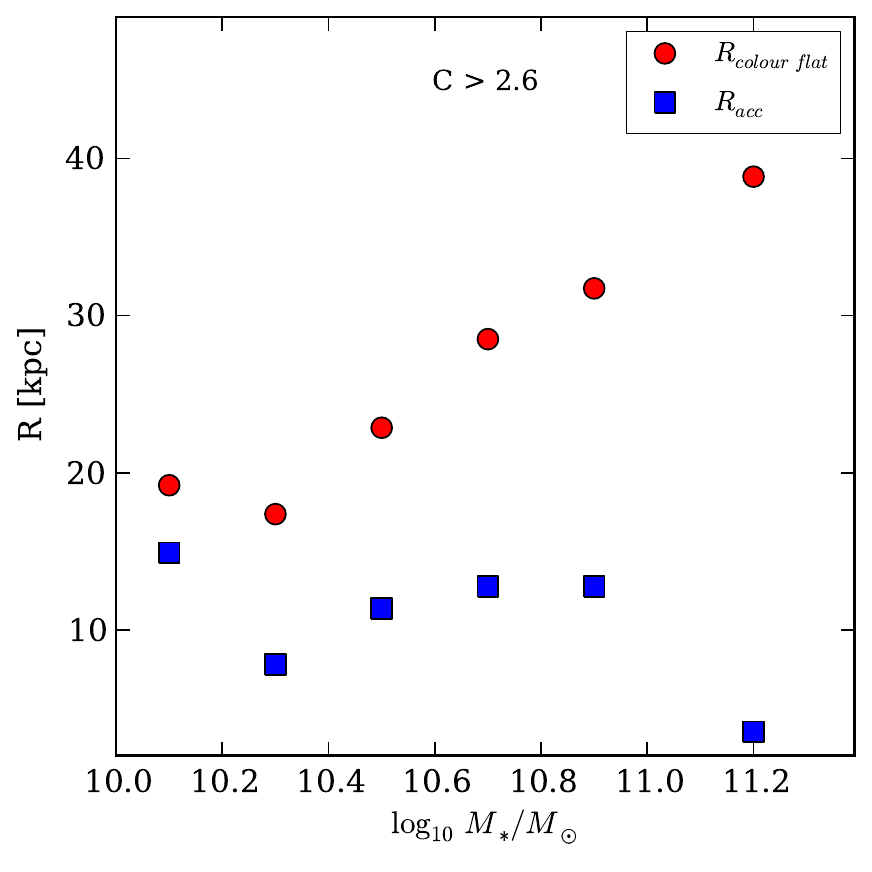}
  \caption{(a and b) The radius at which the accreted component begins to dominate over the in-situ component ($R_{acc}$) for low concentration  
    and high concentration galaxies as a function of stellar mass (blue squares). Also shown is the radius at which there is a minimum in the $g$-$r$ 
    colour profiles ($R_{colour\,min}$) for low concentration galaxies and the radius at which the $g$-$r$ colour profile ($R_{colour\,flat}$) 
    flattens for high concentration as a function of stellar mass (red circles).}
  \label{fig:colour_acc}
\end{figure}

Our modelling allows us to identify a radius at which the outer component begins to dominate 
the integrated stellar light ($R_{acc}$). 
In Figure \ref{fig:colour_acc}, the blue squares indicate this radius as
a function of $M_*$ for low concentration galaxies.  As can be seen, this radius
{\em decreases} as a function of the stellar mass of the galaxy from
$\sim 50$ kpc for galaxies with stellar masses of a few times $10^{10} M_{\odot}$ to
$\sim 30$ kpc for galaxies with $M_* \sim 10^{11} M_{\odot}$. For comparison, we also 
we compare $R_{acc}$ with the radius at which the minimum occurs in the $g$-$r$ colour 
profiles of low concentration galaxies ($R_{colour\,min}$; see Figure \ref{fig:highlowconc}).
The radius at which the outer material begins to dominate is much larger than
the radius at which the minimum in the colour profile occurs. 
This accords well with suggestions in the literature that this minimum in the $g$-$r$ 
colour profile is associated with the break radius in disk galaxies \citep{Bakos}. 

Also in Figure \ref{fig:colour_acc}, we compare the radius at which the outer material begins to dominate
with the radius at which the $g$-$r$ colour profile flattens for high concentration galaxies. The radius at which
the $g$-$r$ colour profile flattens increases as a function of stellar mass from
$\sim 20$ kpc for galaxies with stellar masses of a few times $10^{10} M_{\odot}$ to
$\sim 40$ kpc for galaxies with $M_* \sim 10^{11} M_{\odot}$. The radius at which the outer
material begins to dominate is comparatively smaller and decreases as a function of stellar mass. 
For the highest stellar bin, this radius approaches close to the centre of the galaxy indicating 
that the outer accreted material is spread all over the galaxy. 

\section{Summary} 
\label{sec:summary}
In this work, we have shown that stacking $g$ and $r$ band mosaics of similar galaxies 
allows us to derive reliable surface brightness profiles
upto a depth of $\mu_r \sim 32 \,\mathrm{mag\,arcsec}^{-2}$.  
We study surface brightness, ellipticity and $g$-$r$ colour profiles  as a function of stellar mass and galaxy type.   
We perform fits to the stacked images using multi-component S\'{e}rsic models. This enables us to
estimate the fraction of the stellar light/mass in the outermost component, which
we hypothesize to be built up from accreted stellar material, and 
to set constraints on theories for the formation of stellar haloes 
through hierarchical merging.

The main results of this paper can be summarized as follows.
\begin{enumerate}
\item The fraction of accreted stellar material increases with stellar mass.
At fixed mass, the fraction of accreted material is higher in early-type than in late-type galaxies.
\item The stellar haloes of high concentration galaxies ($C>2.6$) tend to 
be more elliptical than those of low concentration galaxies ($C<2.6$). The ellipticity of the
outer stellar halo increases strongly with stellar mass for high concentration galaxies,
and more weakly with stellar mass for low concentration galaxies.
\item Because we stack galaxies that are nearly face-on, we are only able to 
probe the colour of the outer accreted component in
high concentration galaxies. In these systems, the  $g$-$r$ colour of the outer   
halo light is bluer than than the centre of the galaxy
and is an increasing function of stellar mass.
\item We find that a single-S\'{e}rsic profile cannot fit the entire two-dimensional 
surface brightness distribution of any of our stacked images . Multi-component models  are 
needed to model the excess light in the outer parts of the galaxy, 
especially between  $\mu_r \sim 28 - 32 \,\mathrm{mag\,arcsec}^{-2}$, and to account
for the radial dependence of the  ellipticity of the light distribution.
\item Double-S\'{e}rsic profiles adequately model the surface brightness 
distributions of high concentration galaxies ($C>2.6$), while triple-S\'{e}rsic profiles 
are often needed to model the surface brightness profile of low concentration galaxies ($C<2.6$).
\item Using the fraction of light in the outer component of our models as a measure of
the fraction of the total stellar mass composed of accreted stellar material, we find that this fraction 
is an increasing function of stellar mass. At fixed stellar mass, 
it is also a function of concentration. 
For high concentration galaxies, the fraction of accreted stellar light rises from  $30\%$ to $70\%$, while for 
low concentration galaxies the fraction of stellar light rises 
from  $2\%$ to $25\%$ for galaxies in the stellar mass range $10^{10.0} \msun$ to  $10^{11.4} \msun$. 
\end{enumerate}

\section{Discussion}
\label{sec:discussion}
We have attempted to characterise the stellar halo of galaxies through 
modelling their surface brightness. It is the depth, 
the large dynamic range and the two-dimensional shape information (ellipticity) 
of our surface brightness profiles which enables us to recognise 
deviations from a single component profile and to model successfully 
the stellar halo of our galaxy stacks out to 100 kpc with two or three components. 

An important outcome is that a single S\'{e}rsic component cannot fit the surface brightness 
profiles of high concentration galaxies over a large dynamic 
range in radius and surface brightness, but can only fit the inner parts of galaxies. The inability of a single 
S\'{e}rsic to fit the two-dimensional surface brightness profile of 
galaxies has also been confirmed by the studies of \cite{Bernardi}, \cite{Simard} 
and \cite{Lackner}. Multi-component models are needed to model the full two-dimensional 
surface brightness profiles of galaxies. 
We have demonstrated that it is both the average shape 
of the surface brightness profile and the radial variation in ellipticity
of the light in a galaxy stacks that constrain such models.

For high concentration galaxies, the effective radius of the outer component is twice as large 
as the effective radius of the inner component. For low concentration 
galaxies, the effective radius of the outer component is much larger than the inner components.
For high concentration galaxies, the luminosity of the outer component is 
a significant fraction of the total luminosity of the galaxy and ranges from $30\%$ to $70\%$. 
It also  dominates over a large radial range of the galaxy. On the other 
hand, in low concentration galaxies, the outer component occupies a smaller fraction
(from $2\%$ to $25\%$) and is only dominant at radii larger than  $20-30$ kpc. In both cases, 
the fraction of light in the outer component increases with stellar mass
(see the red line in the top plots of Figure \ref{fig:light-fraction_lc} and Figure \ref{fig:light-fraction_uc}). 

We propose in this work that the fraction of light in the outer component provides 
a measure of the amount of accreted stellar light in the galaxy. While a direct one-to-one 
correspondence between the fraction of light in the outer component and 
the fraction of accreted stellar light cannot be directly proven, the trends in the fraction of light in the 
outer component agree  qualitatively with the trends of the accreted light fraction 
as a function of mass and galaxy-type in the particle-tagging models of  \citep{Cooper}.  
Interestingly, the rate of increase of accreted stellar mass increases 
dramatically above $M_{*} \sim 10^{10.6} \msun$. Interestingly, this corresponds to the stellar mass
where galaxies transition from blue/star-forming to red/passive systems \citep{Kauffmann03b}.   
A significant jump in the accreted mass fraction may be most simply explained by
\emph{in-situ} growth of the galaxy being terminated  by feedback processes, such as energy
injection from relativistic jets produced by black holes in massive galaxies \citep{Croton}.
In the two stage model of massive galaxy formation proposed by  \cite{Oser}, an early, rapid \emph{in-situ} star 
formation period is followed by a late merger-dominated period. In the later phase, galaxies tend to grow predominately through minor mergers. 
We note that the particle tagging  models of  \cite{Cooper} are directly tied to semi-analytic models
that include AGN feedback prescription, and thus  also include quenching of \emph{in-situ} growth of galaxies through cooling
and star formation. In future work, we intend to undertake a detailed comparison with these models. 

Measuring the ellipticity of the outer stellar halo of galaxy also provides us with
hints about the formation processes for the stellar halo. A high ellipticity
is likely to imply that satellite systems are preferentially
accreted along the major axis of the main galaxy  \citep{Tal}.  
The variance in the outer stellar halo profile between different galaxies can be predicted 
from our surface brightness profiles. This variance results from the fact that similar
galaxies can have stellar haloes with very different masses, sizes and shapes. 
The physical origin of this variance as predicted by the $\Lambda$CDM models, is that galaxies of the same
mass have had a range of merger histories, resulting in different accreted stellar mass fractions. 
This has also been clearly  demonstrated using  particle-tagging techniques
on the Aquarius haloes \citep{Cooper10}, which show very large halo-to-halo differences. 

We also note that the integrated surface brightness of the galaxy, including the stellar halo,
includes considerably more light that measured by the SDSS \texttt{model} 
and \texttt{cModel} magnitudes. For example, for high concentration galaxies 
in the stellar mass range $10^{11.0} \msun < M_{*} < 10^{11.4} \msun$, there is about 50\% more light 
contained in the stellar halo at surface brightnesses greater than $\mu_r \sim 24.5 \,\mathrm{mag\,arcsec}^{-2}$. 
This implies that there is considerably more stellar material in the 
galaxy that one might infer from the SDSS photometry. The stellar masses defined by the MPA-JHU 
catalogue and used in this work are only used to define the stellar mass bins, and are  
systematically less than the true stellar mass of the galaxy. This will also be the subject of future work.

\section*{Acknowledgements}
Funding for SDSS-III has been provided by the Alfred P. Sloan Foundation, the
Participating Institutions, the National Science Foundation, and the U.S.
Department of Energy Office of Science. The SDSS-III web site is
\url{http://www.sdss3.org/}.

SDSS-III is managed by the Astrophysical Research Consortium for the
Participating Institutions of the SDSS-III Collaboration including the
University of Arizona, the Brazilian Participation Group, Brookhaven National
Laboratory, University of Cambridge, Carnegie Mellon University, University of
Florida, the French Participation Group, the German Participation Group,
Harvard University, the Instituto de Astrofisica de Canarias, the Michigan
State/Notre Dame/JINA Participation Group, Johns Hopkins University, Lawrence
Berkeley National Laboratory, Max Planck Institute for Astrophysics, Max Planck
Institute for Extraterrestrial Physics, New Mexico State University, New York
University, Ohio State University, Pennsylvania State University, University of
Portsmouth, Princeton University, the Spanish Participation Group, University
of Tokyo, University of Utah, Vanderbilt University, University of Virginia,
University of Washington, and Yale University.

\appendix
\section{The Amount of Light Missed}
\label{appendix:light}
The masking procedure we have employed is far from perfect. Contamination may arise from the incomplete masking of unresolved sources. 
An estimate of the amount of light missed as a function of environment can be made by creating mock galaxy images from an appropriate 
Schechter luminosity function for that environment.
For the purpose of estimating how much of unresolved sources is not masked out in our field environment, we generate 1000 realistic 
mock galaxy $r$-band images resembling the field environment of our Sample by using a fixed single S\'{e}rsic model for the main central 
galaxy and the parameters of the $r$-band Schechter luminosity function of \cite{Blanton2003} for the galaxy environment. Each galaxy image
was convolved with the SDSS $r$-band PSF. In addition, Poisson noise was added to each image.

\begin{figure}
  \includegraphics[width = \linewidth]{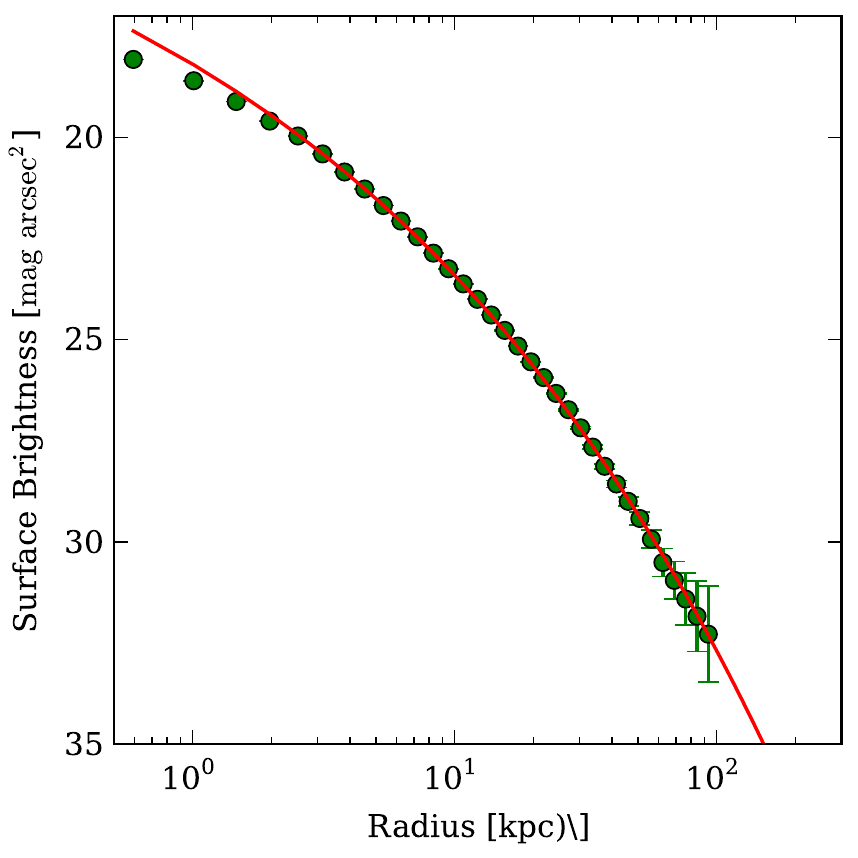}
  \caption{Recovered Luminosity Profiles from the mock images. The red line is the initial model convolved with the $r$-band PSF. 
  }
  \label{fig:lum_compare}
\end{figure}

After subjecting these mocked images to our masking/stacking procedure outlined in the paper, we try to recover the surface brightness profile 
of the central galaxy. We find that we recover surprisingly well the surface brightness profile over a large range of the galaxy as seen in 
\ref{fig:lum_compare}. PSF effects come into play at the centre of the galaxy, while the profile in the faint outer parts depends on the accuracy 
of the background subtraction. 

Our recovery of surface brightness profile can be attributed to a number of factors: First of all, the relatively low density environment of field galaxies help 
in the masking procedure. Secondly, multiple runs of SExtractor help us to mask out most of the over-lapping galaxies. Thirdly, the 
percentile cuts we have used in the stacking procedure helps us to deal with failures in the masking procedure especially close 
to the main galaxy.

\section{Measurement of the Outer Slope}
\label{appendix:slope}
To measure the outer slope ($\mathbf{m}$) of the surface brightness profile, we consider a hierarchical Bayesian methodology that takes into consideration
measurement errors and the intrinsic scatter in the slope $\sigma$ \citep{Kelly}. Following Equation \ref{equation:bayes}, we can write the likelihood
for each measurement $y_{i}$ with measurement error $\delta y_{i}$ as:
\begin{equation}
p(y_{i} \mid \theta) = \frac{1}{\sqrt{2 \pi ( \delta y_{i}^2 + \sigma^2)}} \exp \left\{ -\frac{1}{2}\frac{[y_{i}  - E(y_{i} \mid \theta)]^2}{ \delta y_{i}^2 + \sigma^2} \right\},
\label{equation:likelihood2}
\end{equation}
where $E(y_{i} \mid \theta) = 10^{m  \log x_{i} + c}$. 

Following \cite{Kelly}, we use uniform priors in $m$ ($-10:10$), $c$ ($-100:100$) and $\sigma^2$ ($10^{-8}:1$). We calculate the posterior PDF of 
each parameter using {\sc Multinest}. For the final parameters, we report the maximum of the posterior PDF. The uncertainty in the reported 
parameter is calculated from the variance of the posterior PDF.

\label{lastpage}

\end{document}